\documentclass{elsart}

\usepackage{amsmath,amssymb,graphicx,psfrag,bm,subfigure}

\def\lsim{\mathrel{\lower2.5pt\vbox{\lineskip=0pt\baselineskip=0pt
\hbox{$<$}\hbox{$\sim$}}}}
\def\gsim{\mathrel{\lower2.5pt\vbox{\lineskip=0pt\baselineskip=0pt
\hbox{$>$}\hbox{$\sim$}}}}

\newcommand{\be}{\begin{equation}}
\newcommand{\ee}{\end{equation}}
\newcommand{\gev}{\text{ GeV}}
\newcommand{\mev}{\text{ MeV}}
\newcommand{\Ls}{\Lambda(1405)}
\newcommand{\ba}{\begin{eqnarray}}
\newcommand{\ea}{\end{eqnarray}}

\newcommand{\re}{\text{Re}}
\newcommand{\im}{\text{Im}}

\newcommand{\largeN}[1]{\mbox{\rm ``$#1$''}}

\newcommand{\qmax}{q_{\text{max}}}

\newcommand{\cal}{\mathcal}

\begin{document}

\begin{frontmatter}



\title{On the nature of the $\Lambda(1405)$ and $\Lambda(1670)$ \\
from their $N_c$ behavior in chiral dynamics}


\author[YITP,Murcia]{Luis~Roca\corauthref{cor}},
\corauth[cor]{Corresponding author.}
\author[Munich,YITP]{Tetsuo~Hyodo},
\author[YITP]{Daisuke~Jido}

\address[YITP]{Yukawa Institute for Theoretical Physics, 
Kyoto University, \\
Kyoto 606--8502, Japan}
\address[Murcia]{Departamento de F\'{\i}sica. Universidad de Murcia.
E-30071, Murcia. Spain}
\address[Munich]{Physik-Department, Technische Universit\"at M\"unchen, \\
D-85747 Garching, Germany}

\begin{abstract}
We study the behavior with the number of colors ($N_c$) of the 
$\Lambda(1405)$ and $\Lambda(1670)$ resonances obtained dynamically 
within the chiral unitary approach. The leading order meson-baryon 
interaction, used as the kernel of the unitarization procedure, manifests a 
nontrivial $N_c$ dependence of the flavor $SU(3)$ representation for 
baryons. As a consequence, the $SU(3)$ singlet (or $\bar{K}N$) component of the
$\Lambda(1405)$ states remains bound in the large $N_c$ limit, while the 
other components dissolve into the continuum. Introducing explicit $SU(3)$ 
breaking, we obtain the $N_c$ dependence of the excitation energy, masses and 
widths of the physical $\Lambda(1405)$ and $\Lambda(1670)$ resonance. 
The $N_c$ behavior of the decay width is found to be different 
from the general counting rule for a $qqq$ state, indicating the dynamical 
origin of these resonances.
\end{abstract}

\end{frontmatter}

\section{Introduction}

In hadron physics, one of the most active issues which have continuously 
attracted the attention is the explanation of the inner structure of the 
hadronic resonances. In the last years, it has been suggested that, for many 
mesonic and baryonic resonances, the multi-quark and/or the  hadronic 
components prevail over the simple mesonic  $q\bar q$ and  baryonic $qqq$ 
states. 
A remarkable example is the case of the light scalar mesons ($\sigma$,
$f_0(980)$, $a_0(980)$, $\kappa(900)$), which have been investigated in
four-quark picture~\cite{Jaffe:1976ig}, in lattice QCD~\cite{Alford:2000mm,Kunihiro:2003yj,McNeile:2007fu} and with special success in chiral 
dynamics~\cite{Dobado:1996ps,Oller:1997ti,Oller:1998hw,Oller:1999zr}. In the
baryonic sector the $\Lambda(1405)$ resonances has been well described in 
coupled-channel meson-baryon 
scattering~\cite{Dalitz:1960du,Dalitz:1967fp},
in a constituent quark model~\cite{Isgur:1978xj},
in lattice QCD~\cite{Melnitchouk:2002eg,Nemoto:2003ft,Lee:2005mr,Ishii:2007ym}, and
in dynamical approach with chiral symmetry~\cite{Kaiser:1995eg,Oset:1998it,Oller:2000fj,Lutz:2001yb}. Given the importance to unveil the nature of the 
hadronic spectrum in order to further understand the dynamics of the strong 
interaction, methods to clarify the internal structure of the hadrons are 
called for. This is one of the main aims of the present work, in the 
baryonic sector, following the lines presented in an exploratory 
study~\cite{Hyodo:2007np}.
 
It is well known the success of QCD as the theory for the strong 
interaction. However, the application of QCD to the low and intermediate 
hadronic spectrum is not straightforward, due to the confinement property. 
It is at this point where the importance of 
effective theories
emerges, like 
chiral perturbation theory
(ChPT)~\cite{Weinberg:1979kz,Gasser:1983yg,Gasser:1984gg}. 
Despite the success of ChPT to explain a vast 
amount of hadronic phenomenology at low energies, 
ChPT has a very important 
limitation. The applicable energy range of ChPT is 
typically up to the lowest resonance in a particular channel because ChPT is
based on perturbative expansions of momenta,
which can never reproduce the singularity 
associated to a resonance pole. Furthermore the perturbative expansions 
violate unitarity of scattering matrix at a certain kinematical scale. Hence, in order to 
construct scattering amplitudes able to make predictions and to explain the 
interesting hadronic region around $1 \gev$, we should rely upon 
non-perturbative methods. An outstanding success among many efforts in this 
direction has been the so called chiral unitary 
approach~\cite{Dobado:1996ps,Oller:1997ti,Oller:1998hw,Oller:1999zr,Kaiser:1995eg,Oset:1998it,Oller:2000fj,Lutz:2001yb}. Its power and beauty 
stem from the fact that it is able to reproduce a vast amount of hadronic 
phenomenology up to energies even beyond $1\gev$  with the only input of the
lowest orders ChPT Lagrangians, the requirement that the amplitudes must 
fulfill unitarity in coupled channels and the exploitation of the analytic 
properties of the scattering amplitude. The chiral unitary approach provides
not only scattering amplitudes to explain hadronic interactions but also 
exploits the possibility to obtain dynamically many hadronic resonances not 
initially present in the ChPT Lagrangian. This is the case, for instance, of
the low lying scalar mesons~\cite{Dobado:1996ps,Oller:1997ti,Oller:1998hw,Oller:1999zr,Kaiser:1998fi,Markushin:2000fa} which appear from the 
interaction of pseudoscalar mesons, or most of the lightest axial-vector 
resonances from the interaction of a pseudoscalar and a vector 
meson~\cite{Lutz:2003fm,Roca:2005nm}. Another case of successful application
of these chiral unitary techniques is the interaction of mesons with 
baryons~\cite{Kaiser:1995eg,Oset:1998it,Oller:2000fj,Lutz:2001yb,Kaiser:1995cy,Kaiser:1996js,Krippa:1998us,Nieves:1998hp,Nacher:1999vg,Oset:2001cn,Inoue:2001ip,Jido:2002yz,Jido:2002zk}. These remarkable successes in a variety of channels can be 
understood that the leading order chiral interaction is determined model 
independently~\cite{Weinberg:1966kf,Tomozawa:1966jm}, which is the 
driving force to generate the resonances~\cite{Hyodo:2006yk,Hyodo:2006kg}

Particularly interesting is the discovery that the $\Lambda(1405)$ resonance
is actually a superposition of two states~\cite{Fink:1989uk,Oller:2000fj,Jido:2002yz,Jido:2003cb}. By looking at the strangeness $S=-1$, isospin 
$I=0$, $s$-wave meson-baryon scattering amplitude, two poles are found in 
the second Riemann sheet at the positions $1390-i66\mev$ and 
$1426-i16$~\cite{Oset:2001cn,Jido:2002yz,Jido:2003cb} (similar positions are found in 
Refs.~\cite{Hyodo:2002pk,Hyodo:2003qa,Borasoy:2005ie,Oller:2005ig,Oller:2006jw,Borasoy:2006sr}). 
These poles strongly couple to the $\pi \Sigma$ and $\bar K N$ channels, respectively~\cite{Jido:2003cb}.
The physical origin of this interesting 
structure is attributed to the attractive interaction in \textit{both} the 
$\bar{K}N$ and the $\pi\Sigma$ channels, constrained by chiral low energy 
theorem~\cite{Hyodo:2007jq}. The phenomenological consequences of the 
two-pole structure of the $\Lambda(1405)$ have recently been tested by the 
reaction $K^-p\rightarrow\pi^0\pi^0\Sigma^0$~\cite{Prakhov:2004an} which was
theoretically demonstrated in Ref.~\cite{Magas:2005vu}, by the 
$pp \to p K^+ \Lambda(1405)$ reaction in Refs.~\cite{Zychor:2007gf,Geng:2007vm}, by the reaction $\pi^-p\to K^0\pi\Sigma$ in 
Ref.~\cite{Hyodo:2003jw}, by the radiative decay into $\gamma\Lambda$ and 
$\gamma\Sigma^0$ in Ref.~\cite{Geng:2007hz}, and by $K^*$ photoproduction in
Ref.~\cite{Hyodo:2004vt}. The $\Ls$ has been typically one of the most 
poorly understood baryons; as is well known, its low mass has been quite 
difficult to understand in naive constituent quark 
models~\cite{Isgur:1978xj},
and the spectra shape of the $\Lambda(1405)$ was found to be
incompatible with a Breit-Wigner 
shape~\cite{Hemingway:1984pz,Dalitz:1991sq}.
 The two-pole picture shed new light on the 
structure of the $\Ls$, which is shown to have a large impact for the study 
of the  $\bar{K}N$ interaction~\cite{Hyodo:2007jq}. This is eventually 
important for the study of the kaonic nuclei~\cite{Akaishi:2002bg,Yamazaki:2007cs,Dote:2008in} and kaon condensation in neutron 
stars~\cite{Kaplan:1986yq}. 
The key issue is the resonance position of the $\Lambda(1405)$ in
the $\bar K N$ channel. The nominal $\Lambda(1405)$ observed in the
$\pi \Sigma$ final interaction has a resonance peak around 
1405~MeV, while the two-pole picture suggests that the
$\Lambda(1405)$ having strong coupling to $\bar KN$ is located
at a higher energy, around 1425~MeV. The difference looks very small
but, in the $\bar KN$ bound state picture of the $\Lambda(1405)$,
the important quantity is the bounding energy measured from the
$\bar KN$ threshold. These two values are physically quite
different.  
Thus, it is desired to understand the structure 
of the $\Lambda(1405)$ resonance.

The expansion in powers of the inverse of the number of colors, $1/N_{c}$, 
is an analytic, well established and widely used approach to QCD valid 
for the whole energy region, and 
it enables us to investigate the qualitative 
features of hadrons~\cite{Hooft:1974jz,Witten:1979kh,Witten:1980sp}. In the 
last years the $N_c$ dependence of the poles associated to resonances within
the chiral unitary approach has shown up as a powerful tool to study the 
internal structure of particular mesonic resonances~\cite{Pelaez:2003dy,Pelaez:2006nj,Pelaez:2004xp,newaxialsNc}. Since the $1/N_{c}$ expansion is 
valid for the whole energy region, it can be applied to the low energy 
hadron scattering and hence it provides a connection between the quark 
language and the effective theories with hadronic degrees of freedom, like 
the chiral unitary approach. The known $N_c$ behavior of $q\bar{q}$ mesons 
enable us to discriminate such component from the others
(for example, see Ref.~\cite{Jaffe:2007id}).
 For instance, the 
$N_c$ behavior of the $\sigma$ meson is totally at odds with a predominant 
$q\bar q$ nature, while the $\rho$ meson follows clearly the expected 
$q\bar q$ behavior~\cite{Pelaez:2003dy}. A similar conclusion regarding the 
dynamical nature studying the $N_c$ behavior of the resonance poles was 
found for the case of most of the low lying axial-vector 
resonances~\cite{newaxialsNc}.

The situation is by far more complicated in the baryon sector due to the
nontrivial $N_c$ dependence of the leading order meson-baryon interaction 
found in Refs.~\cite{Hyodo:2006yk,Hyodo:2006kg}. It is usually considered 
that the leading order chiral interaction scales as ${\cal O}(N_c^{-1})$,
due to the factor $1/f^2$. However, the flavor representation of the baryons
changes with $N_c$, when the number of flavors is larger than
two~\cite{Karl:1985qy,Dulinski:1987er,Dulinski:1988yh}. As a consequence, 
the group theoretical factor of  the leading order chiral meson-baryon 
interaction shows a nontrivial $N_c$ dependence, and for some cases, it
scales as ${\cal O}(1)$. This would provide a different $N_c$ behavior
to what would typically be expected from dynamically generated resonances, 
that is, the widths of the resonances go to infinity for large $N_c$ 
(see, for instance, Ref.~\cite{Jido:2003cb}).
As we will see, the fact that the poles
go to infinite width for large $N_c$ is true for the most components, 
however, the flavor singlet component shows a non-trivial behavior: it becomes 
bound in the large $N_c$ limit. This can also be understood in connection 
with the kaon bound state approach to the Skyrmion~\cite{Callan:1985hy,Itzhaki:2003nr}.

At the same time, the behavior with $N_c$ 
of several resonance properties can provide information
about the quark 
structure of the resonance. The general $N_c$ counting rule for ordinary
$qqq$ baryons indicates the scaling of the decay width as  $\Gamma_R\sim
{\cal O}(1)$, the mass $M_R~\sim {\cal O}(N_c)$ and the  excitation energy
$\Delta E\equiv M_R-M_B-m\sim{\cal O}(1)$ with $M_B$  ($m$)
 the ground-state
baryon (meson) mass~\cite{Hooft:1974jz,Witten:1979kh,Witten:1980sp,Goity:2004pw,Cohen:2003fv}. Hence any significant deviation 
from these behaviors would indicate that other non-$qqq$ components (like 
hadronic molecules) dominate over the $qqq$ contribution in the wave 
function of the baryonic resonance. The $N_c$ behavior of the baryon 
resonances was studied in Ref.~\cite{GarciaRecio:2006wb} in the SU(6) 
symmetric limit. A recent work \cite{Hyodo:2007np} studied for the first
time the $N_c$ behavior of the physical baryon resonances with flavor
symmetry breaking effects, focusing on the $\Ls$ resonance. In the present 
paper, we extend  the work of Ref.~\cite{Hyodo:2007np} by carefully
explaining the details of the analysis, investigating the $N_c$ behavior of more 
resonance properties and studying in addition
 the $\Lambda(1670)$ resonance which 
also appears dynamically in the strangeness $S=-1$, isospin $I=0$, 
meson-baryon scattering amplitude.

In section~\ref{sec:uchpt} we summarize the formalism of the chiral unitary
approach in order to obtain the $S=-1$ meson-baryon unitarized scattering 
amplitudes. In section~\ref{sec:arbNc} we explain the extension of the model
to arbitrary $N_c$ paying special attention to the lowest order meson-baryon
interaction and to the regularization scheme of the unitary bubble. 
We will start explaining our results, with the consideration, in 
section~\ref{sec:largeN}, of the large $N_c$ limit where one would expect a 
bound state for the $SU(3)$ singlet and $\bar{K}N$ channels. In 
section~\ref{sec:physical} the explicit $SU(3)$ breaking is taken into 
account and we study the $N_c$ behavior of the poles and the resonance 
parameters in order to look for discrepancies from the general counting 
rules for $qqq$ states. We will finish by summarizing the results and 
conclusions.

\section{The chiral unitary approach for $\Lambda(1405)$ and $\Lambda(1670)$}
\label{sec:uchpt}

Detailed explanations of the formalism for the construction of the
meson-baryon unitarized amplitude can be found in Refs.~\cite{Oset:1998it,Oller:2000fj,Hyodo:2006kg,Oset:2001cn,Jido:2003cb,Hyodo:2003qa}. In the 
following we give a brief review of the formalism for the sake of completeness. 
The lowest order chiral Lagrangian for the interaction of the flavor octet of Nambu-Goldstone bosons with the octet of the low lying $1/2^+$ baryons is given 
by~\cite{Bernard:1995dp}: 
\begin{equation}
    \mathcal{L}=\frac{1}{4f^2}\langle\bar{B}i\gamma^\mu
    [\Phi\partial_\mu\Phi-\partial_\mu\Phi\Phi,B]\rangle, 
    \label{eq:Lag}
\end{equation}
with $\Phi$ ($B$) the usual $SU(3)$ matrices containing the octets of
pseudoscalars (baryons) and $f$ the meson decay constant. Equation~\eqref{eq:Lag}
provides the tree level transition amplitudes which, for 
a center of mass energy $W$ and after projecting over $s$-wave, are
\begin{equation}
    V_{ij}(W)=-C_{ij}\frac{1}{4f^2}(2W-M_i-M_j)
    \left(\frac{M_i+E_i}{2M_i}\right)^{1/2}
    \left(\frac{M_j+E_j}{2M_j}\right)^{1/2}, 
    \label{eq:WT}
\end{equation} 
with  $E_i$ ($M_i$) the energies (masses) of the baryons of the $i$-th 
channel and $C_{ij}$ coefficients group-theoretically given by 
\begin{equation} C_{ij}^{I} =\begin{pmatrix} 3 &
-\sqrt{\frac{3}{2}} & \frac{3}{\sqrt{2}} & 0 \\ & 4 & 0 &
\sqrt{\frac{3}{2}} \\ &   & 0 & -\frac{3}{\sqrt{2}} \\ &   &   & 3
\end{pmatrix} 
\label{eq:couplingI},
\end{equation}
for $I=0$ and $S=-1$ with $C_{ij}=C_{ji}$.
The $i$ and $j$ subindices represent the different meson-baryon channels 
which in this case are, in isospin basis, $\bar K N$, $\pi\Sigma$, 
$\eta\Lambda$ and $K\Xi$. The superscript $I$ in Eq.~(\ref{eq:couplingI}) 
means that the matrix is given in isospin basis, in contrast to the $SU(3)$ 
basis that will be introduced below. Equation~(\ref{eq:WT}) is known as the
Weinberg-Tomozawa term derived using current algebra~\cite{Weinberg:1966kf,Tomozawa:1966jm}.

The key point of the chiral unitary approach is the implementation of 
unitarity in coupled channels. Based on the $N/D$ method~\cite{Oller:1999zr,Oller:2000fj,Hyodo:2003qa}, the coupled-channel scattering amplitude 
$T_{ij}$ is given by the matrix equation
\begin{equation}
   T=[1-VG]^{-1}V ,
   \label{eq:BS}
\end{equation}
where $V_{ij}$ is the interaction kernel of Eq.~(\ref{eq:WT}) and the 
function $G_{i}$, or unitary bubble, is given by the dispersion integral of 
the two-body phase space 
$\rho_i(s)=2M_{i}\sqrt{(s-s_i^+)(s-s_i^-)}/(8\pi s)$ in a diagonal matrix 
form by
\begin{align}
   G_i(W)
   &=-\tilde{a}_i(s_0)
   -\frac{s-s_0}{2\pi}
   \int_{s_i^{+}}^{\infty}ds^{\prime}
   \frac{\rho_i(s^{\prime})}{(s^{\prime}-s)(s^{\prime}-s_0)}
   \label{eq:loop_s} , \\
   &s=W^{2} , \quad s_i^{\pm}=(m_i\pm M_{i})^2,
   \nonumber
\end{align}
where $s_0$ is the subtraction point, $\tilde{a}_i(s_0)$ the subtraction 
constant and $m_i$ is the mass of the meson of the channel $i$. This 
expression corresponds to the meson-baryon loop function. We will explain 
the different regularization procedures adopted for the $G$ function, in 
connection with the $N_c$ scaling, in section~\ref{subsec:regularization}.

The matrix elements $T_{ij}$ of Eq.~(\ref{eq:BS}) provide the 
$M_j B_j\to M_i B_i$ scattering amplitudes which satisfy elastic unitarity 
in coupled channels. The presence of a resonance in a given partial wave 
amplitude must be identified as poles of the scattering matrix in unphysical
Riemann sheets. If these poles are not very far from the real axis the pole 
position, $s_R\equiv W_R^2$, is related to the mass, $M_R$, and width, 
$\Gamma_R$, of the resonance by $W_R=(M_R\pm i\Gamma_R/2)$. The analytic 
structure of the scattering amplitude is determined by the loop function 
$G$. The $G$ function in the second Riemann sheet (II) can be obtained
from the one in the first sheet (I) by \cite{Roca:2005nm}
\begin{equation}
G_i^{II}(W)=G_i^{I}(W)+iM_i\frac{\bar{q}_i}{2\pi W},
\label{eq:GII}
\end{equation}
\noindent
with $\bar{q}_i$ the center of mass meson or baryon momentum,
\begin{equation}
    \bar{q}_i = \frac{\sqrt{(s-(M_i-m_i)^2)(s-(M_i+m_i)^2)}}{2W} ,
    \nonumber
\end{equation}
with $\im(\bar{q})>0$. When looking for poles we will use $G_j^I(W)$ for 
$\re(W)<m_j+M_j$ and $G_j^{II}(W)$ for $\re(W)>m_j+M_j$. This prescription 
gives the pole positions closer to those of the corresponding Breit-Wigner 
forms on the real axis.
 
Close to the pole position the amplitude can be approximated by its Laurent 
expansion where the dominant term is given by
\be
    T_{ij}(W)
    =
    \frac{g_i\,g_j}{W-M_R+i\Gamma_R/2},
    \label{eq:Tpole}
\ee
for an $s$-wave resonance. Consequently the residue of $T_{ij}$ at the pole 
position gives $g_ig_j$, where $g_i$ is the effective coupling of the 
dynamically generated resonance to the $i$-th channel.

It is also interesting and relevant for the forthcoming
discussion to obtain the couplings of the resonances to the 
states labeled by the $SU(3)$ irreducible representations. 
The scattering amplitude of the octet meson and octet baryon can be
decomposed into the following irreducible representations:
\begin{equation}
  {\bm 8} \otimes {\bm 8} = {\bm 1} \oplus {\bm 8} \oplus {\bm 8}^{\prime}
  \oplus \bm{10} \oplus \bm{\overline{10}} \oplus \bm{27}
\end{equation}
Among the above representations, only
  ${\bm 1}$, ${\bm 8}$, ${\bm 8}^{\prime}$
and  $\bm{27}$ are relevant for $I=0$.
The couplings in the $SU(3)$ basis can be obtained 
by transforming the amplitude from the isospin 
basis to the $SU(3)$ one by means of the matrix $U$, where 
$U_{i\alpha}=\langle i , \alpha\rangle$, with $\alpha$ labeling the $SU(3)$ 
state. By using the $SU(3)$ Clebsch-Gordan 
coefficients~\cite{deSwart:1963gc,McNamee:1964xq}, the explicit expression
of $U$ for $S=-1$ and $I=0$ is
\be
U_{i\alpha} \equiv 
\left(\begin{array}{cccc}
-\frac{1}{2} & -\frac{1}{\sqrt{10}} &   \frac{1}{\sqrt{2}}  & -\frac{1}{2}\sqrt{\frac{3}{5}} \\
\frac{1}{2}\sqrt{\frac{3}{2}} & -\sqrt{\frac{3}{5}} &   0  & -\frac{1}{2\sqrt{10}} \\
-\frac{1}{2\sqrt{2}} & -\frac{1}{\sqrt{5}} &  0& \frac{3}{2}\sqrt{\frac{3}{10}} \\
\frac{1}{2}&   \frac{1}{\sqrt{10}}  &  \frac{1}{\sqrt{2}}& \frac{3}{2\sqrt{5}} \\
\end{array}
\label{eq:U3}
\right) ,
\ee
where the isospin states are in the order $\bar K N$, $\pi\Sigma$, 
$\eta\Lambda$, $K\Xi$ and the $SU(3)$ ones in the order $\bm{1}$, $\bm{8}$, 
$\bm{8}'$, $\bm{27}$. A matrix quantity $X^I$ in isospin basis can be 
transformed into that in $SU(3)$ basis by
\be
X^{SU(3)}=U^\dagger\,X^{I}\,U .
\label{eq:basistrans}
\ee
The matrix $X$ can be the amplitude $T$ and the interaction $V$. If $X$ is 
an $SU(3)$ symmetric quantity, its expression in $SU(3)$ basis should be a 
diagonal matrix. For instance, the $C$ matrix in $SU(3)$ basis is given by
\begin{equation}
    C_{\alpha\beta}^{SU(3)}
    =\begin{pmatrix}
        6  \ \ \ & 0\ \ \  & 0\ \ \  & 0  \\
                 & 3\ \ \  & 0\ \ \  & 0  \\
                 &         & 3\ \ \  & 0  \\
                 &         &         & -2
     \end{pmatrix}
    \label{eq:couplingSU3}.
\end{equation}
The physical scattering amplitude contains the $SU(3)$ breaking, so that the 
$T_{\alpha\beta}^{SU(3)}$ is no longer a diagonal matrix. However, 
the resonance poles are independent of the channels as seen in Eq.~\eqref{eq:Tpole}.
All the channel dependence is summarized in the residues, or the couplings
of the resonance to the channels. 
Applying 
Eq.~\eqref{eq:basistrans} to the resonance amplitude~\eqref{eq:Tpole}, we 
obtain the coupling strength of the resonance in the $SU(3)$ basis as
\be
(g_1, g_8, g_{8'}, g_{27})=(g_{\bar K N},g_{\pi\Sigma},
g_{\eta\Lambda}, g_{K\Xi})\, U ,
\label{eq:gU}
\ee
which is also valid for the physical scattering with $SU(3)$ breaking. 

The signs of the coefficients in Eq.~(\ref{eq:couplingSU3}) imply that the 
meson-baryon interaction in the $SU(3)$ symmetric case is attractive for the
singlet and the two octets and repulsive for the 27-plet. Actually, when 
looking for poles, it was found in Ref.~\cite{Jido:2003cb} in the $SU(3)$ 
symmetric case one pole for the singlet and two degenerate poles in the real
axis for the octets. As $SU(3)$ is gradually broken by using different 
masses within each multiplet, the $SU(3)$ states mix and two branches emerge
from the initial octet pole position and one from the singlet pole. The 
branches finish in the physical mass limit situation with two poles very 
close to the nominal $\Ls$ resonance and another one corresponds to the 
$\Lambda(1670)$ resonance~\cite{Oller:2000fj,Jido:2003cb}. In tables~1 and 5
of Ref.~\cite{Jido:2003cb} the different poles and their corresponding
couplings to the different isospin and $SU(3)$ meson-baryon states are 
shown. We can see in those tables that the lower mass pole ($z_1$ in the 
following) of the $\Ls$ has the wider width and couples dominantly to 
$\pi\Sigma$, while the higher mass pole ($z_2$) has the narrower width and 
couples dominantly to $\bar K N$. This is understood also by the attractive 
interactions in $\pi\Sigma$ and $\bar{K}N $ channels in isospin 
basis~\cite{Hyodo:2007jq}. By looking at the $SU(3)$ couplings it can be 
seen that the $z_1$ pole has retained dominantly the singlet nature it had 
in the $SU(3)$ symmetric situation but it has an admixture with the octet. 
The $z_2$ pole of the $\Ls$ has, in the physical situation, dominant 
component of the singlet but it also has a strong contribution from the 
octet states.

\section{Extension of the model to arbitrary $N_{c}$
}
\label{sec:arbNc}

In this section, we discuss how to extrapolate the present model shown
in the previous section to arbitrary number of colors, $N_{c}$.
Generally it is hard to calculate exact values of the parameters in the effective
Lagrangian directly from QCD, while it is relatively easy to obtain the $N_{c}$
dependence of the parameters in a model independent way. To extend our model
to arbitrary $N_{c}$, we need to know the $N_{c}$  dependence for the ingredients 
of the present formulation appearing in Eqs.~(\ref{eq:WT}) and (\ref{eq:BS}), that
is, the masses and decay constants for the pseudoscalar mesons $m_i$
and $f$, the baryon masses $M_i$, the $C_{ij}$ coefficients and the 
renormalization procedure of the $G$ functions. The $N_{c}$ dependence of 
the mesonic quantities is well-known. 
The meson mass and decay constant scale as ${\cal O}(1)$ and ${\cal
O}(N^{1/2}_{c})$, respectively, in the leading $N_{c}$ expansion.
Here we assume that the meson mass and decay constant in arbitrary
$N_{c}$ are given by $m_i(N_c) = m_i $ and $f(N_c)=f_0 \sqrt{N_c/3}$
with $f_0$ being the value at $N_c=3$. The $N_c$ dependence of the
other quantities related to baryons, that is baryon mass, the
$C_{ij}$ coefficient and the  regularization of the $G$ function,
will be explained in detail below. 

An important point for the $N_c$ scaling of baryons is the assignment
of the flavor  $SU(3)$ representation to the baryon in arbitrary
$N_{c}$. Since, in the $N_{c}$ world, baryons are composed by $N_{c}$
quarks, the flavor contents of the baryon are not trivially given. In
the meson case,  since the number of quarks in a meson does not
change, the flavor of the meson remains the same as the $N_{c}=3$.
For baryons, in principle we have  three ways to extend the flavor representation
of the baryon \cite{Dulinski:1988yh}. In this work, we mainly use the
standard extension which keeps the spin, isospin and strangeness of
the baryon as they are in $N_{c}=3$. This is suitable for discussing
the flavor $SU(3)$ breaking. In the standard extension, the baryon
belonging the $[p,q]$ representation in the tensor notation
extrapolates as
\begin{equation}
    [p,q] \to 
    \left[
    p,q+\frac{N_c-3}{2}\right]
    \label{eq:standard} 
\end{equation}
in arbitrary $N_{c}$. 
In Sec.~\ref{sec:otherNc}, we briefly discuss the other extensions. 
We call also the flavor representation in arbitrary $N_{c}$ for the baryon belonging to 
the $\bm{R}$ representation in $N_{c}=3$ by $\largeN{\bm{R}}$~\cite{Karl:1985qy,Dulinski:1987er,Dulinski:1988yh}.

\subsection{Baryon masses}

The general counting rule for a baryon made up of $N_c$ quarks establishes
that the mass scales as $M\sim \mathcal{O}(N_c)$~\cite{Witten:1979kh}.
The mass splitting due to the flavor $SU(3)$ breaking 
appears from the order of  ${\cal O}(1)$ in the $1/N_{c}$ expansion~\cite{Jenkins:1995td}. 
Thus, we assume that the baryon masses in arbitrary $N_{c}$ are given by
\begin{equation}
   M_{i}=M_0\frac{{N_c}}{3}+ \delta_{i} \label{eq:NcBaryonMass} 
\end{equation}
with $M_{0}=1151$ MeV and the flavor symmetry breaking $\delta_{i}$
being  $\delta_{N}= -212$, $\delta_{\Lambda}=-35$, $\delta_{\Sigma}=42$ and $\delta_{\Xi}=167$ in units of MeV. The $M_{0}$ is the averaged value of the observed 
octet baryon masses, since the other mass splittings than the flavor symmetry breaking
appear from ${\cal O}(1/N_{c})$~\cite{Dashen:1993as,Dashen:1993ac}.
The $SU(3)$ breaking parameters for each octet baryon are 
fixed so that Eq.~(\ref{eq:NcBaryonMass}) reproduces the observed masses
at $N_{c}=3$. With these extensions of the meson and 
baryon masses to arbitrary $N_{c}$, the values of thresholds for the meson-baryon scattering is also extended to arbitrary $N_{c}$ up to ${\cal O}(1)$. 

\subsection{Coupling strengths}

For the $N_c$ dependence of the coupling strengths $C_{ij}$, let us start 
with the strengths in the $SU(3)$ basis. The channel
labeled by the $\bm R$ irreducible representation at $N_{c}=3$ 
belongs to the $\largeN{\bm R}$ representation in arbitrary $N_{c}$, which is given explicitly by Eq.~(\ref{eq:standard}) for the standard extension. 
Once the representation of the channel is fixed, the coupling strength can
be obtained in a diagonal matrix of which the elements are determined 
only by a group theoretical argument~\cite{Hyodo:2006yk,Hyodo:2006kg}:
\begin{align}
     C_{\alpha\beta}^{SU(3)}(N_c)
     &=\begin{pmatrix}
        \frac{9}{2}+\frac{N_c}{2} & 0 & 0 & 0  \\
          & 3 & 0 & 0  \\
          &   & 3 & 0  \\
          &   &   & -\frac{1}{2}-\frac{N_c}{2} 
     \end{pmatrix}
     \label{eq:couplingSU3Nc} .
\end{align}
The channels are in the order
$\largeN{\bm{1}}$, $\largeN{\bm{8}}$, $\largeN{\bm{8'}}$, 
$\largeN{\bm{27}}$. At $N_c=3$, this is reduced to 
Eq.~\eqref{eq:couplingSU3}. The coupling strength in the isospin basis can 
be obtained by the Clebsch-Gordan coefficients  with $N_c$ 
dependence~\cite{Cohen:2004ki}. Taking into account the phase factor 
correctly, we obtain the matrix $U(N_c)$ as
\begin{align}
    &U_{i\alpha}(N_c) \equiv \nonumber \\
    &\begin{pmatrix}
	-\sqrt{\frac{N_c-1}{N_c+5}} & -\sqrt{\frac{6(N_c-1)}{D}} &   2(N_c+2)\sqrt{\frac{6}{D(N_c+7)}}  & -2\sqrt{\frac{3}{(N_c+7)(N_c+5)}} \\
	\sqrt{\frac{3}{N_c+5}} & \frac{-3(N_c+1)}{\sqrt{2D}} &   (N_c-3)\sqrt{\frac{N_c-1}{2D(N_c+7)}}  & -\sqrt{\frac{N_c-1}{(N_c+7)(N_c+5)}} \\
	-\sqrt{\frac{3(N_c-1)}{(N_c+5)(N_c+3)}} & -(N_c+9)\sqrt{\frac{N_c-1}{2D(N_c+3)}} &   -3(N_c-3)\sqrt{\frac{N_c+3}{2D(N_c+7)}}  & 3\sqrt{\frac{N_c+3}{(N_c+7)(N_c+5)}} \\
	2\sqrt{\frac{3}{(N_c+5)(N_c+3)}}&   6\sqrt{\frac{2}{D(N_c+3)}}  & 5\sqrt{\frac{2(N_c+3)(N_c-1)}{D(N_c+7)}}& \sqrt{\frac{(N_c+3)(N_c+1)}{(N_c+7)(N_c+5)}} \\
     \end{pmatrix}
    \label{eq:UNc} ,
\end{align}
with $D=5N_c^2+22N_c+9$. This corresponds to Eq.~\eqref{eq:U3} at $N_c=3$.
The coupling strengths in the isospin basis are given through
Eq.~\eqref{eq:basistrans} by
\begin{equation}
    C_{ij}^I(N_c)=
    \begin{pmatrix}
        \frac{1}{2} (3+{N_c})
	& -\frac{\sqrt{3}}{2} \sqrt{-1+{N_c}}
	& \frac{\sqrt{3}}{2} \sqrt{3+{N_c}}
	& 0   \\
	& 4
	& 0
	& \frac{\sqrt{3+N_c}}{2} \\ 
	& 
	& 0
	& -\frac{3}{2} \sqrt{-1+{N_c}}\\
	& 
	& 
	&  \frac{1}{2} (9-{N_c}) \\
     \end{pmatrix}
    \label{eq:couplingINc} ,
\end{equation}
with $ C_{ij}^I= C_{ji}^I$. 
The channels are in the same order as before, which means the counterparts
in arbitrary $N_{c}$ to 
$\bar{K}N$, $\pi\Sigma$, $\eta\Lambda$, and $K\Xi$ at $N_{c}=3$
\footnote{The baryons here should be understood to belong the $\largeN{\bm 8}$ and
have different charge and hypercharge from the $N_{c}=3$ case.}. 
Note that strengths in the diagonal $\bar{K}N$ 
and $K\Xi$ channels are $\mathcal{O}(N_c)$, and the negative $N_c$ 
dependence in the $K\Xi$ channel changes the sign of the interaction from 
attraction to repulsion for $N_c>9$. On the other hand, the off-diagonal 
elements (and diagonal $\pi\Sigma$ and $\eta\Lambda$ ones) are 
not more than $\mathcal{O}(\sqrt{N_c})$, 
so that any transitions among these channels vanish faster than $1/\sqrt{N_c}$ 
in large $N_{c}$ together with an extra $N_{c}$ dependence coming from $1/f^2$ factor in the interaction~\eqref{eq:WT}. 
This means that the meson-baryon 
scattering for these channels becomes essentially a set of single-channel 
problems in large $N_c$ limit, even with the $SU(3)$ breaking in the meson 
and baryon masses. This point 
will be important in the discussion of the large $N_c$ limit in 
section~\ref{subsec:largeNbound}

\subsection{Regularization procedure}
\label{subsec:regularization}

The $G_i$ function in Eq.~(\ref{eq:BS}) can also be interpreted as the loop 
function of a meson and a baryon 
\be
    G_i(W)
    = i\int\frac{d^{4}q}{(2\pi)^{4}}
    \frac{2M_i}{(P-q)^{2}-M_i^{2}+i\epsilon}
    \frac{1}{q^{2}-m_i^{2}+i\epsilon} 
    \label{eq:loop}  ,
\ee
\noindent 
with  the incoming four-momentum $P=(W,0,0,0)$ in the center of mass frame. 
Since this loop function diverges logarithmically, 
an appropriate regularization procedure is necessary to proceed the calculation,
such as the three-momentum cut-off and the dimensional regularization. 
The regularization brings parameters which cannot be fixed within the scattering
theory. These parameters have turned out to be very important for the structure of the 
dynamical generated resonances as shown in Refs.~\cite{Hyodo:2007jk,Hyodo:2008xr}. Thus, we discuss several regularization schemes, paying strong attention to 
the $N_{c}$ dependence in the regularization parameters. 


For an interpretation of the scale of the ultraviolet cutoff, the 
momentum cutoff scheme is suitable. Adopting the three-momentum 
cutoff $q_{\text{max}}$ for $|\vec q|$, we can regularize the loop function as
\ba
    G_i^{3d}(W)
    &=&\frac{2M_{i}}{(4\pi)^{2}}
    \Biggl\{\ln\frac{m_iM_i}{\qmax^{2}}
    +\frac{\Delta_i}{s}\ln\frac{M_i(1+\xi_i^m)}{m_i(1+\xi_i^M)}
    -\ln[(1+\xi_i^m)(1+\xi_i^M)]
    \nonumber \\
    &&+\frac{\bar{q}_i}{W}
    [\text{Ln}^m_{i,+}(s)+\text{Ln}^M_{i,+}(s)
    -\text{Ln}^m_{i,-}(s)-\text{Ln}^M_{i,-}(s)]\Biggl\} \nonumber ,
\ea
with
\begin{align}
    \Delta_i 
    =&  M_i^{2}-m_i^{2} , \nonumber \\
    \text{Ln}^{m}_{i,\pm}(s) 
    =& \ln[\pm s\mp \Delta
    +2\bar{q}_iW\xi^{m}_i] ,
    \quad 
    \text{Ln}^{M}_{i,\pm}(s) 
    = \ln[\pm s\pm \Delta
    +2\bar{q}_iW\xi^{M}_i],
    \nonumber \\
    \xi^{m}_i
    =& \sqrt{1+\frac{m_i^2}{\qmax^2}},
    \quad 
    \xi^{M}_i
    = \sqrt{1+\frac{M_i^2}{\qmax^2}} .
    \nonumber 
\end{align}
The value $q_{\text{max}}\simeq 630\mev$ was used in Ref.~\cite{Oset:1998it}
to reproduce the meson-baryon scattering in $S=-1$ channel. The fact that 
the order of magnitude of the cutoff is about $1\gev$ can be understood from
the point of view of the effective theory in the following way. We can 
consider two possible scenarios for the origin of the numerical value of the
cutoff~\cite{newaxialsNc}: i) it corresponds to the 
scale of the spontaneous chiral symmetry breaking $q_{\text{max}}\sim 4\pi f\sim 1\gev$. ii)  it corresponds to the mass of a heavier $qqq$ state 
integrated out in order to construct the effective theory. These 
interpretations of the cutoff provide a natural $N_c$ scaling of 
$q_{\text{max}}$, 
If the i) 
scenario determines the energy scale of the cutoff, then the $N_{c}$ 
scaling of the cutoff should be $q_{\text{max}}\sim{\cal O}(\sqrt{N_c})$ since 
$f\sim{\cal O}(\sqrt{N_c})$. Therefore, a natural integral cutoff should 
scale as $\sqrt{N_c}$ but no faster, otherwise the loop momentum could get 
values larger than the scale of the effective theory. We call this scheme as ``scaling cutoff''. In the ii) 
scenario the cutoff would scale as $q_{\text{max}}\sim{\cal O}(1)$ since
the energy difference between ground state baryons and excited baryons scale
as ${\cal O}(1)$. We refer to this scheme as ``unscaling cutoff''. In 
anyway, the options of the $N_c$ dependence on the cut-off parameters
should  be only viewed as indicative and representative of possible $N_c$ 
dependence of the cutoff.
The difference between the results obtained with the dependence is
an estimation of the uncertainty from this source. 

The dimensional regularization scheme provides the 
loop function whose analytic structure is consistent with the dispersion 
integral~\eqref{eq:loop_s}:
\ba
    G^{\text{dim}}_i(W)&=&\frac{2M_{i}}{(4\pi)^{2}}
    \Biggl\{a_i(\mu)+\ln\frac{m_iM_i}{\mu^{2}}
    +\frac{\Delta_i}{s}\ln\frac{M_i}{m_i}
    \nonumber \\
    &&+\frac{\bar{q}_i}{W}
    [\text{Ln}_{i,++}(s)+\text{Ln}_{i,+-}(s)
    -\text{Ln}_{i,-+}(s)-\text{Ln}_{i,--}(s)]\Biggl\} \nonumber ,
\ea
with
\begin{align}
    \text{Ln}_{i,+\pm}(s) 
    &=\ln[  s\pm \Delta_i+2\bar{q}_iW ],
    \quad
    \text{Ln}_{i,-\pm}(s) 
    =\ln[ - s\pm \Delta_i+2\bar{q}_iW ] .
    \nonumber
\end{align}
There is one degree of freedom of regularization, corresponding to the 
cutoff $\qmax$ in the three-momentum cutoff scheme. For this degree of 
freedom we introduce the parameter $\mu_m$ at which 
$G^{\text{dim}}_i(W=\mu_m) = 0$ is required. In the $SU(3)$ symmetric limit,
we choose this scale at baryon mass $\mu_m = M_i$. For the physical 
scattering case, the scale $\mu_m$ is taken to be $M_{\Lambda}$ for all 
channels as in Ref.~\cite{Lutz:2001yb}, which gives a reasonable description
of the $\bar{K}N$ scattering. The scale $\mu_m$ is regarded as the matching 
scale of the full amplitude $T_{ij}$ to the interaction kernel $V_{ij}$, and
can be used to study the origin of the resonances as explained in detail in 
Refs.~\cite{Hyodo:2007jk,Hyodo:2008xr}. 
In this regularization, we take the matching scale $\mu_m$ at arbitrary $N_{c}$ 
as the $\Lambda$ mass $M_\Lambda$ given in 
Eq.~(\ref{eq:NcBaryonMass}).

In this section, we have discussed three different scalings for the regularization 
procedure: two (scaling and unscaling) for the cutoff and one for the 
dimensional regularization methods. In the analysis of physical resonances, 
we will consider all three possibilities to see uncertainties of the theory. 
As we will find later, however, 
all three procedures give qualitatively similar results,  leading to the 
same conclusions.

\section{Analysis in the large $N_c$ limit}
\label{sec:largeN}
%

\subsection{Bound state in the large $N_c$ limit}
\label{subsec:largeNbound}

In this section, we consider  the meson-baryon 
scattering and the possibility of having a bound state in the large $N_{c}$ limit. 
Let us start with the 
problem in $SU(3)$ symmetric limit for simplicity. In this case, there is no
channel coupling, and we can follow the argument given in 
Refs.~\cite{Hyodo:2006yk,Hyodo:2006kg} for single-channel scattering. The 
mass of the target hadron is simply given by $M_T(N_c)=M_0 N_c/3$ where 
$M_0$ is the value at $N_c=3$. Adopting the dimensional regularization 
scheme, we study the existence of a bound state through the critical 
coupling $C_{\text{crit}}$ introduced in Refs.~\cite{Hyodo:2006yk,Hyodo:2006kg}. The critical coupling is calculated for arbitrary $N_c$ as
\begin{equation}
    C_{\text{crit}}(N_c)= \frac{2[f(N_c)]^2 }{m[-G(M_{T}(N_c)+m)]} .
    \label{eq:Ccrit}
\end{equation}
If the coupling strength~\eqref{eq:couplingSU3Nc} is 
larger than this critical value, a bound state is generated
 in the scattering amplitude.
At $N_c=3$, the critical coupling is $C_{\text{crit}}\sim 2.4$,
so the attractive forces in $\bm{1}$, $\bm{8}$ and
$\bm{8}^{\prime}$ generate three bound states.

\begin{figure}[tbp]
    \centering
    \includegraphics[width=.45\linewidth,clip]{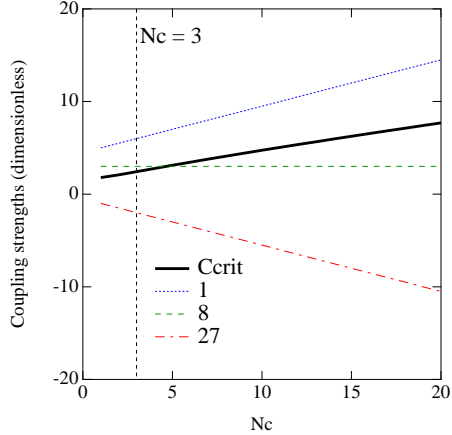}
    \caption{\label{fig:critical}
    $N_c$ dependence of the coupling strengths $C_{\alpha}$ with 
    $\alpha = \largeN{\bm{1}}$, $\largeN{\bm{8}}$, and $\largeN{\bm{27}}$
    (the dotted, dashed, and dash-dotted lines, respectively) together with the 
    critical coupling strength (solid line). The vertical dashed line represents
    $N_c=3$.}
\end{figure}

We plot $C_{\text{crit}}(N_c)$ in Fig.~\ref{fig:critical}, with $M_0=1151$ 
MeV and $f_0=93$ MeV together with the coupling strengths relevant
for $S=-1$ and $I=0$ channel. As seen in the figure, $C_{\text{crit}}(N_c)$ 
increases as $N_{c}$ increases. 
It can be proved that $C_{\text{crit}}(N_c)$ does not increase
faster than $N_c/2$, which is the $N_{c}$ dependence of  
the coupling strength for the $\largeN{\bm{1}}$ channel in the large $N_{c}$ limit. 
Indeed, in ref.~\cite{Hyodo:2006kg} a explicit expression for 
$G(M_{T}(N_c)+m)$ is given 
(see eqs.(7) and (21) of ref.~\cite{Hyodo:2006kg}),
from where it can be easily shown that, in the large $N_c$ limit, 
$G(M_{T}+m)$ goes as

\begin{equation}
   G(M_{T}(N_c)+m) \longrightarrow 
    \frac{m}{4\pi^{2}}\ln \frac{M_{T}(N_c)}{m}.
\end{equation}
Therefore, since $M_T(N_c)$ and $f(N_c)$ increase as ${\mathcal
  O}(N_c)$ and ${\mathcal O}(N_c^{1/2})$, respectively, the critical
  coupling scales as ${\mathcal O}(N_c^{n<1})$.
This is slower than ${\cal O}(N_{c})$. Thus the positive linear $N_c$ dependence of the $C_{ij}$ coefficient always generate a bound state in the large $N_c$ limit. On the 
other hand, the attraction in $\largeN{\bm{8}}$ channel becomes smaller than
$C_{\text{crit}}(N_c)$ at larger value of $N_c$. In the large $N_c$ limit, 
therefore, there is only one bound state in the singlet channel, instead of three, found at
$N_c=3$.

As we have noted in Eq.~\eqref{eq:couplingINc}, the off-diagonal couplings of
the meson-baryon interaction in the isospin basis vanish 
in the large $N_c$ limit. This 
means that, even with $SU(3)$ breaking, the scattering in the
isospin basis behaves as a single-channel problem at sufficiently large $N_c$.
The coupling strength of $\bar{K}N$ channel in the large $N_c$ limit is the 
same as the one of the singlet channel in the $SU(3)$ basis. Therefore, by 
following the same argument as the $SU(3)$ symmetric case, 
we conclude that there is one bound 
state in the $\bar{K}N$ channel in the large $N_{c}$ limit. It was found in 
Ref.~\cite{Hyodo:2007jq} that the $\bar{K}N$ interaction develops a bound 
state at $N_c=3$, when the transition to other channels are switched off. 
Thus, as in the $SU(3)$ singlet channel, the $\bar{K}N$ bound state found at
$N_c=3$ remains in the large $N_c$ limit in contrast
to the mesonic resonances, while the other states, such as a resonance in $\pi\Sigma$ 
channel, disappears in the large $N_{c}$ limit.

In this section, we have found that the meson-baryon state remains bound 
in the large $N_c$ 
limit, both in the $SU(3)$ singlet channel and in the $\bar{K}N$ channel. It
is interesting to consider the relation of this result to the kaon bound 
state approach to the Skyrmion~\cite{Callan:1985hy,Itzhaki:2003nr}. In this 
picture based on $1/N_c$ expansion, the $\Lambda(1405)$ is described as 
one bound state of an antikaon and a nucleon. In the chiral unitary 
approach, the $\Lambda(1405)$ is described as the two poles generated by the
attractive interaction in $\bar{K}N$ and $\pi\Sigma$ channels. But taking 
the large $N_c$ limit, only one of them survives as a bound state in the 
$\bar{K}N$ channel, which may correspond to the bound state found in the 
Skyrmion approach. This ensures the correct behavior of our amplitude at 
large $N_c$ limit, while at the same time, it also suggests the importance 
of the $\pi\Sigma$ interaction which is the subleading effect of the $1/N_c$
expansion, when one consider the physical system with $N_c=3$.

\subsection{Other $N_c$ extensions}
\label{sec:otherNc}
Thus far we have used the standard large $N_c$ extension
for the baryon representations given in Eq.~(\ref{eq:standard}). 
The merit of this generalization is that spin, isospin and strangeness of
the baryon are the same as those at $N_c=3$, while the baryon has 
different charge and hypercharge from those at $N_{c}=3$. 
It is known that there are two other extensions of the baryon flavor in 
arbitrary $N_{c}$~\cite{Dulinski:1988yh}:
\begin{align}
    [p,q] \to &
    \left[
    p+\frac{N_c-3}{3},q+\frac{N_c-3}{3}\right]
    \label{eq:ext2} ,\\
    [p,q] \to &
    \left[
    p+N_c-3,q\right]
    \label{eq:ext3} .
\end{align}
These extensions have some advantages, but the baryons constructed
in these  ways have unphysical strangeness and spin (see
Ref.~\cite{Piesciuk:2007xq}  for detail). With these extensions,
the general expressions of the coupling strengths of the WT term
are given in  Table~\ref{tbl:largeNgeneral}. It can be seen that
the strengths of the  exotic channels $[p+1,q+1]$, $[p+2,q-1]$ and
$[p-1,q+2]$ have negative or  constant dependence on $N_c$. In the
context of Refs.~\cite{Hyodo:2006yk,Hyodo:2006kg}, we confirm that
the nonexistence of the attractive interaction in exotic channel
in any generalizations of the flavor representations.

\begin{table}[tbp]
    \centering
    \caption{General expressions of the coupling strengths 
    $C_{\largeN{\alpha},\largeN{T}}$ for the baryon representation $T=[p,q]$
    with three different $N_c$ extensions, Eqs.~\eqref{eq:standard}, 
    \eqref{eq:ext2} and \eqref{eq:ext3}.}
    \begin{tabular}{cccc}
        \hline
        $\alpha$ & 
	\eqref{eq:standard} 
	& \eqref{eq:ext2} 
	& \eqref{eq:ext3}   \\
        \hline
        $[p+1,q+1]$ & $\frac{3-N_c}{2}-p-q$ & $\frac{6-2N_c}{3} -p-q$ 
	& $3-N_c-p-q$  \\
        $[p+2,q-1]$ & $1-p$ & $\frac{6-N_c}{3}-p$
	& $4-N_c+p$  \\
        $[p-1,q+2]$ & $\frac{5-N_c}{2}-q$ & $\frac{6-N_c}{3}-q$ 
	& $1-q$  \\
        $[p,q]$ & 3 & 3 & 3  \\
        $[p,q]$ & 3 & 3 & 3  \\
        $[p+1,q-2]$ & $\frac{3+N_c}{2}+q$ & $\frac{6+N_c}{3}+q$ 
	& $3+q$  \\
        $[p-2,q+1]$ & $3+p$ & $\frac{6+N_c}{3}+p$
	& $N_c+p$  \\
        $[p-1,q-1]$ & $\frac{5+N_c}{2}+p+q$ & $\frac{6+2N_c}{3} +p+q$ 
	& $1+N_c+p+q$  \\
        \hline
    \end{tabular}
    \label{tbl:largeNgeneral}
\end{table}

Turning back to the $S=-1$ and $I=0$ scattering, the relevant channels are 
$\largeN{\alpha} = \largeN{\bm{1}}$, $\largeN{\bm{8}}$, and 
$\largeN{\bm{27}}$ with $\largeN{T}=\largeN{\bm{8}}$. The coupling strengths
of these channels
are given in Table~\ref{tbl:coupling}. As seen in the table, the 
coefficients of the linear $N_c$ dependence for the $\largeN{\bm{1}}$ channel in nonstandard 
extensions~\eqref{eq:ext2} and \eqref{eq:ext3} are $2/3$ and $1$, which are 
larger than that in the standard extension $1/2$. This means that the 
singlet bound state exists in all cases. Thus, the qualitative conclusion
of the previous subsection remains unchanged when we adopt different methods
of the extension of the flavor representations.

\begin{table}[tbp]
    \centering
    \caption{Coupling strengths relevant for $S=-1$ and $I=0$ meson-baryon 
    scattering with $\largeN{T}=\largeN{\bm{8}}$.}
    \begin{tabular}{cccc}
        \hline
        $\largeN{\alpha}$ & \eqref{eq:standard} & \eqref{eq:ext2}
	& \eqref{eq:ext3} \\
        \hline
        $\largeN{\bm{1}}$ & $\frac{9}{2}+\frac{N_c}{2}$ 
	& $4+\frac{2N_c}{3}$ & $3+N_c$  \\
        $\largeN{\bm{8}}$ & 3 & 3 & 3  \\
        $\largeN{\bm{27}}$ & $-\frac{1}{2}-\frac{N_c}{2}$
	& $-\frac{2N_c}{3}$ & $1-N_c$  \\
        \hline
    \end{tabular}
    \label{tbl:coupling}
\end{table}

\section{Structures of resonances and comparison with quark picture}
\label{sec:physical}


In this section we investigate the structure of the resonances obtained in the present
formulation with $SU(3)$ breaking by comparing the quark picture of the resonances.  
The $SU(3)$ breaking effects are implemented
in the masses of the pseudoscalar mesons and baryons, which are fitted 
by the observed masses at $N_{c}=3$. 
The present model successfully reproduces phenomenological properties 
of the $\Lambda(1405)$ and $\Lambda(1670)$ very well at $N_{c}=3$. 
Exploiting the present description of the resonances, we investigate the 
structure of the resonances.
For this purpose, first we show
the $N_{c}$ behaviors of the pole positions (mass and width) and the couplings
to various channels, and then we interpret the $N_{c}$ behaviors in terms
of the inner structure of the resonances by comparing with those expected by the quark picture.

We will not take values of $N_c$ very far 
from $N_c=3$ for the following reason. 
The resonances described by the present formulation
can have some admixture of genuine preexisting $qqq$ components for 
$N_c=3$. These components certainly have different $N_c$ behaviors than the 
dynamical molecule, but we do not consider the $N_{c}$ dependence of such
component. 
Hence, for very large $N_c$ these admixture could be 
different from what the resonances have intrinsically.
For instance, even  if a very small component of the genuine quark state is 
present at $N_c=3$, the genuine quark may show up for 
sufficiently large $N_c$ since the quark states survive in the large $N_c$ 
limit. Thus, in order to study the quark structure of ``physical'' 
resonances, we vary $N_c$ from 3 to 12 in the analysis. 

\subsection{$N_{c}$ behaviors of pole positions and coupling strengths}

We calculate masses and widths of the resonances in the scattering amplitude
given in Eq.~(\ref{eq:BS}) by looking for the poles of the amplitude in the complex energy plane numerically. 
We use the $N_{c}$ dependence of the parameters in this model as already
given in Sec.~\ref{sec:arbNc}. 
With the $SU(3)$ breaking, the scattering amplitude is calculated in coupled channels and each channel has a different threshold.
If the poles appear above the lowest threshold, the pole positions
are expressed by complex number and the poles represent resonances with finite
widths.
As we have discussed in Sec.~\ref{sec:uchpt}, the 
$\Lambda(1405)$ is described by two poles in this approach. 
In addition, this model reproduces the $\Lambda(1670)$ resonance
at the same time.
Thus, at $N_{c}=3$ we have three poles in the complex energy plane. 

\begin{figure}[tbp]
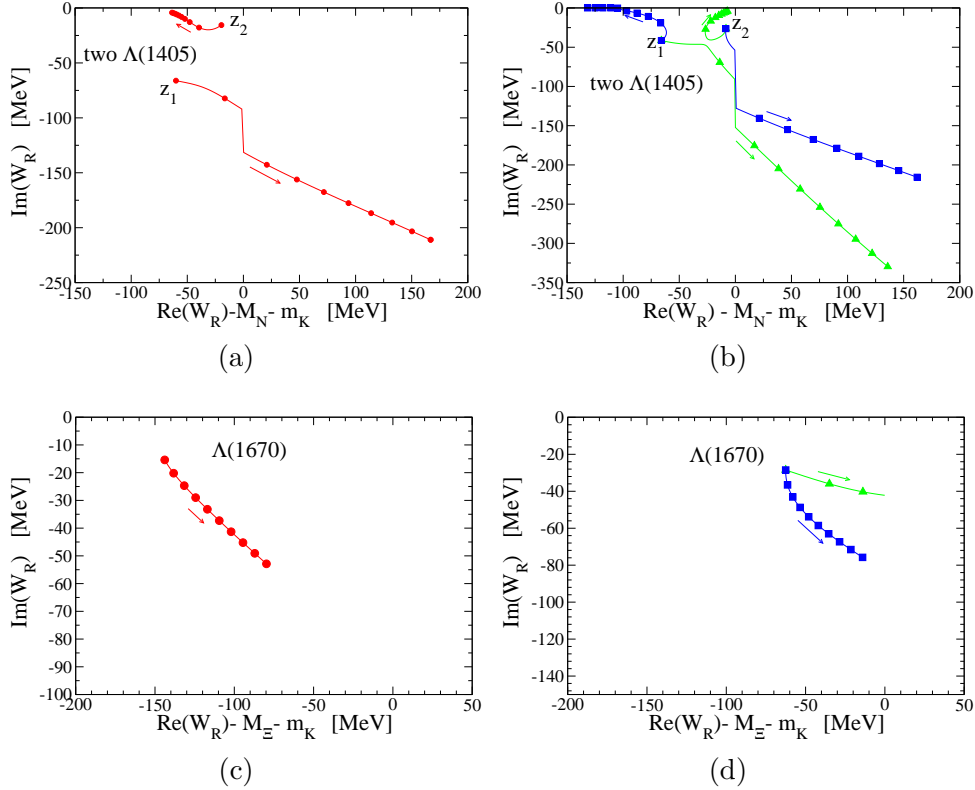

     \centering
     \subfigure[]{
          \label{fig:polesa}
          \includegraphics[width=.45\linewidth]{figure2a.eps}}
     \subfigure[]{
          \label{fig:polesb}
          \includegraphics[width=.45\linewidth]{figure2b.eps}}
   \\ \vspace{0.1cm}
       \subfigure[]{
          \label{fig:polesc}
          \includegraphics[width=.45\linewidth]{figure2c.eps}}
     \subfigure[]{
          \label{fig:polesd}
          \includegraphics[width=.45\linewidth]{figure2d.eps}}
     \caption{Pole positions of the  $s$-wave meson-baryon scattering
     amplitudes with $I=0$ and $S=-1$ as a function of $N_c$ for 
     the three  regularization methods discussed in the text. 
     The horizontal axis denotes the real part of the pole position 
     measured by the thresholds of the $\bar KN$ channel for the $\Lambda(1405)$ 
     and the $K\Xi$ channel for the $\Lambda(1670)$, and the vertical axis expresses 
     the imaginary part of the pole position. 
     The value of $N_{c}$ 
     varies from 3 to 12 as indicated by arrows. 
     The two upper panels correspond
 to the two $\Lambda(1405)$ and the lower panel to the
 $\Lambda(1670)$. The calculation for the plots in the left side, 
 (a) and (c), is done with the dimensional regularization method. 
 The calculation for the plots in the right hand side, (b) and (d), is 
 performed with the cutoff method. The lines with the squares in the
 right panels is for the scaling cut-off, $q_{\text{max}}\sim{\cal O}(\sqrt{N_c})$,
 while the lines with the triangles is for the unscaling cut-off
     $q_{\text{max}}\sim{\cal O}(1)$.}
     \label{fig:poles}
\end{figure}

Calculating the pole positions of the $s$-wave scattering amplitude with $I=0$ and $S=-1$ as functions of $N_{c}$, we obtain the 
$N_{c}$ behavior of the pole position. 
In Fig.~\ref{fig:poles} we show the positions of three poles;
two of them appear in energies of the $\Lambda(1405)$ (upper 
panels) and the other shows up in energies of the $\Lambda(1670)$ (lower panels). 
The horizontal axis in Fig.~\ref{fig:poles}  represents the excitation (or binding) energy 
which is the energy of the resonance measured by the threshold
of the relevant channel for the resonance. We take 
the $\bar KN$ channel  for the $\Lambda(1405)$ and the $K\Xi$ channel for the
$\Lambda(1670)$ as the threshold. Consequently the excitation energies are expressed by $\re(W_R)-M_N-m_K$ for the $\Lambda(1405)$ poles and 
$\re(W_R)-M_\Xi-m_K$ for the $\Lambda(1670)$. The vertical axis expresses
the imaginary part of the pole position.
We also show 
the results for the three different regularization methods of the
$G$ function discussed in Sec.~\ref{subsec:regularization}. 
In the left panels, the results obtained with the dimensional regularization 
are shown, while in the right panels the results with the 
scaling cutoff (square points) and the unscaling cutoff (triangles)
are presented. 
The symbols in the lines are 
placed in steps for $N_c$ of $1$ unit from 3 to 12. 
The discontinuity for 
the two $\Ls$ poles at $\re(W_R)-M_N-m_K=0$ is due to the
threshold effect. When  the pole crosses the $\bar K N$ threshold,
the width suddenly increases 
because the important $\bar K N$ decay channel suddenly opens.
As another threshold effect, it happens that two mathematical poles 
close each other
are obtained around the threshold with a model parameter set. 
This is known as  
``shadow poles" consisting in the presence of two
nearby poles associated to the same resonance if the pole is very
close to a threshold. It is a consequence of unitarity  (see
Ref.~\cite{shadowpoles} for further details).

Regarding the two $\Ls$ poles, one of them approaches the real axis with 
reducing its width as $N_c$ increases. The other $\Ls$ pole moves to higher 
energy region and the imaginary part increases with $N_c$. Namely, one $\Ls$
resonance tends to become a bound state, while the other tends to 
dissipate by moving away from the physical axis. For the $\Lambda(1670)$ 
pole, both the mass and width increase with $N_c$, which indicates the tendency of
dissipation. 
These qualitative findings are independent of the choice of the regularization scheme. 
It is worth mentioning that there is a difference 
among the choices of the regularization methods: in the dimensional 
regularization and unscaling cutoff cases, the pole having higher mass 
at $N_c=3$ ($z_2$) goes to the bound state as $N_{c}$ increases 
and the lower pole dissipates, while in the scaling
cutoff, the behavior is the other way around. 
In fact, it is very sensitive to the value of the meson decay constant
which pole goes to the bound state in larger $N_{c}$. For instance, if 
we use a bit larger decay constant like $f(N_c=3)=1.2\times92.4\mev$
than the original $f(N_c=3)=1.123\times92.4\mev$, 
the three schemes
provide the same behavior, in the sense that the 
higher pole goes to the bound state.  
However it is not relevant physically which pole
goes to the bound state, but it is  important what properties the pole
going to the bound state has. 
As will become clear when we evaluate the
flavor components of the poles,  the properties of the poles are
not dependent on the choice of the  regularization schemes. This
means that they manifest the qualitative  important behavior that
one resonance tends to become a bound  state while the other one
 tends
to dissipate by moving away from the physical  axis.

\begin{figure}[tbp]
     \centering
     \subfigure[]{
          \label{fig:g1g8a}
          \includegraphics[width=.45\linewidth]{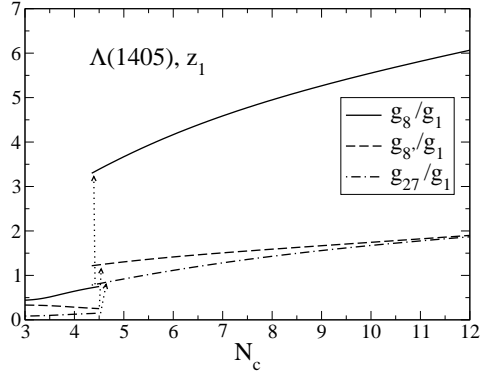}}
   \\ \vspace{0.1cm}
       \subfigure[]{
          \label{fig:g1g8b}
          \includegraphics[width=.45\linewidth]{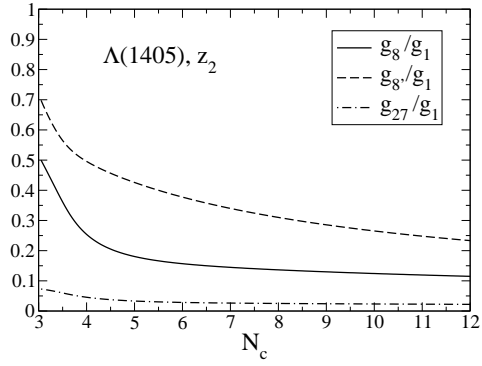}}
   \\ \vspace{0.1cm}
       \subfigure[]{
          \label{fig:g1g8c}
          \includegraphics[width=.45\linewidth]{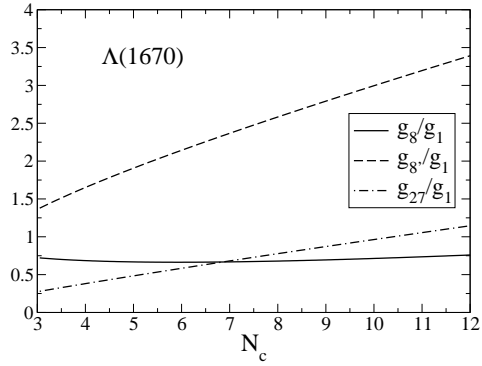}}
     \caption{Couplings of 
 the three poles in the $SU(3)$ basis
 divided by the singlet channel,
 $g_\alpha/g_1$. The calculation is done with the
 dimensional regularization method. The two upper panels correspond
 to the two $\Lambda(1405)$ and the lower panel to the
 $\Lambda(1670)$.} 
 \label{fig:g1g8}
\end{figure}

In order to illustrate further the properties of the dynamically generated 
resonances, it is very interesting to evaluate the $SU(3)$ contents of 
the poles as functions of $N_c$. This also makes clear connection 
with the large $N_c$ analysis in Sec.~\ref{subsec:largeNbound}.
As explained in Sec.~\ref{sec:uchpt}, the $SU(3)$ contents of the poles 
are calculated by the residues of the scattering amplitudes at the pole
positions. The coupling constants in the $SU(3)$ basis are obtained 
in Eq.~(\ref{eq:gU}) by the unitary transformation from the isospin basis.
In Fig.~\ref{fig:g1g8}, we show the $N_c$ dependence of the coupling 
ratio, $g_\alpha/g_1$, for the channel $\alpha$ in absolute value.
The calculations are performed  using the dimensional regularization method. 
In order from the top,  the panels 
represent the lower mass $\Ls$ (at $N_c=3$), the higher mass $\Ls$ 
and the $\Lambda(1670)$ respectively. In the central panel we can see that the 
the higher pole of the $\Ls$ becomes dominantly the singlet flavor-$SU(3)$ state
as $N_{c}$ increases, while the lower $\Ls$ pole, which goes to infinite width, becomes essentially $\bm{8}$ and the $\Lambda(1670)$ one becomes essentially
$\bm{8}'$. This is consistent with
 what we have found in the large $N_{c}$ limit. 

\begin{figure}[tbp]
     \centering
     \subfigure[]{
          \label{fig:gIsospa}
          \includegraphics[width=.45\linewidth]{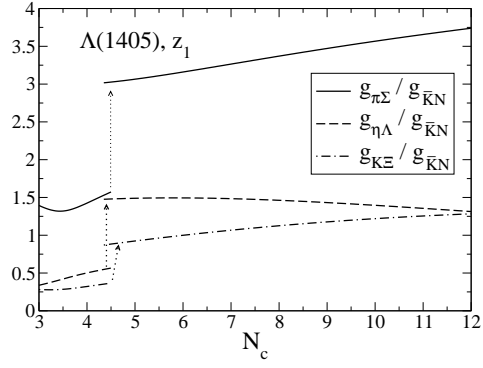}}
   \\ \vspace{0.1cm}
       \subfigure[]{
          \label{fig:gIsospb}
          \includegraphics[width=.45\linewidth]{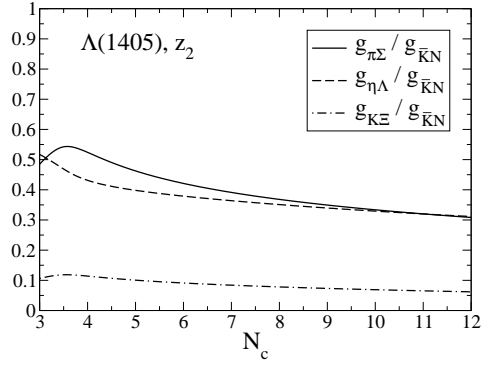}}
   \\ \vspace{0.1cm}
       \subfigure[]{
          \label{fig:gIsospc}
          \includegraphics[width=.45\linewidth]{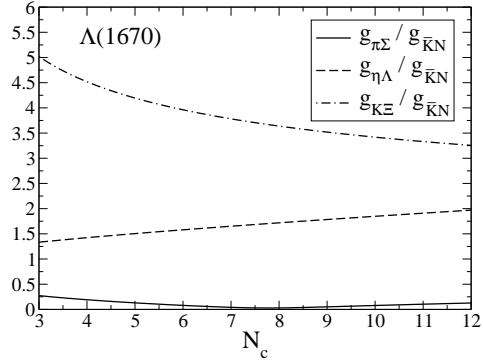}}
     \caption{Couplings of the three poles in the isospin basis
 divided by the $\bar K N$ channel,
 $g_i/g_{\bar K N}$. The calculation is done with the
 dimensional regularization method. The two upper panels correspond
 to the two $\Lambda(1405)$ and the lower panel to the
 $\Lambda(1670)$.} 
 \label{fig:gIsosp}
\end{figure}

In Fig.~\ref{fig:gIsosp} we also show the couplings of the three poles in the isospin basis
by taking their ratios of the coupling to the $\bar K N$ channel for reference,
$|g_i|/|g_{\bar{K}N}|$. We can see in the upper panel that the lower $\Ls$ 
pole, which becomes essentially octet for larger $N_c$, couples dominantly
to the $\pi\Sigma$ state. On the contrary, the higher $\Ls$ pole, shown in the central 
panel, couples dominantly to $\bar K N$. 
The lower panel in Fig.~\ref{fig:gIsosp} shows that the strongest
coupling of the $\Lambda(1670)$ to the $K\Xi$ state gets weaker as
$N_{c}$ increases and the coupling to the $\eta\Lambda$ channel
becomes stronger. 

These analyses of the coupling strengths indicate that the
dominant  component of the resonances associated with the pole
becoming  bound state is  flavor singlet ($\bar{K}N$) in the
$SU(3)$ (isospin) basis, whereas such component in the dissipating
resonance becomes less important.

 It is interesting to evaluate the coupling strengths of the two
$\Lambda(1405)$ poles  with the scaling cutoff method because,
in this
case, as we have already seen above (see Fig.~\ref{fig:polesa} and
the squares in Fig.~\ref{fig:polesb}),
  the movement of the poles is
different from that obtained in the dimensional regularization
scheme, that is, the lower mass pole is the one
 becoming the bound state.
\begin{figure}[tbp]
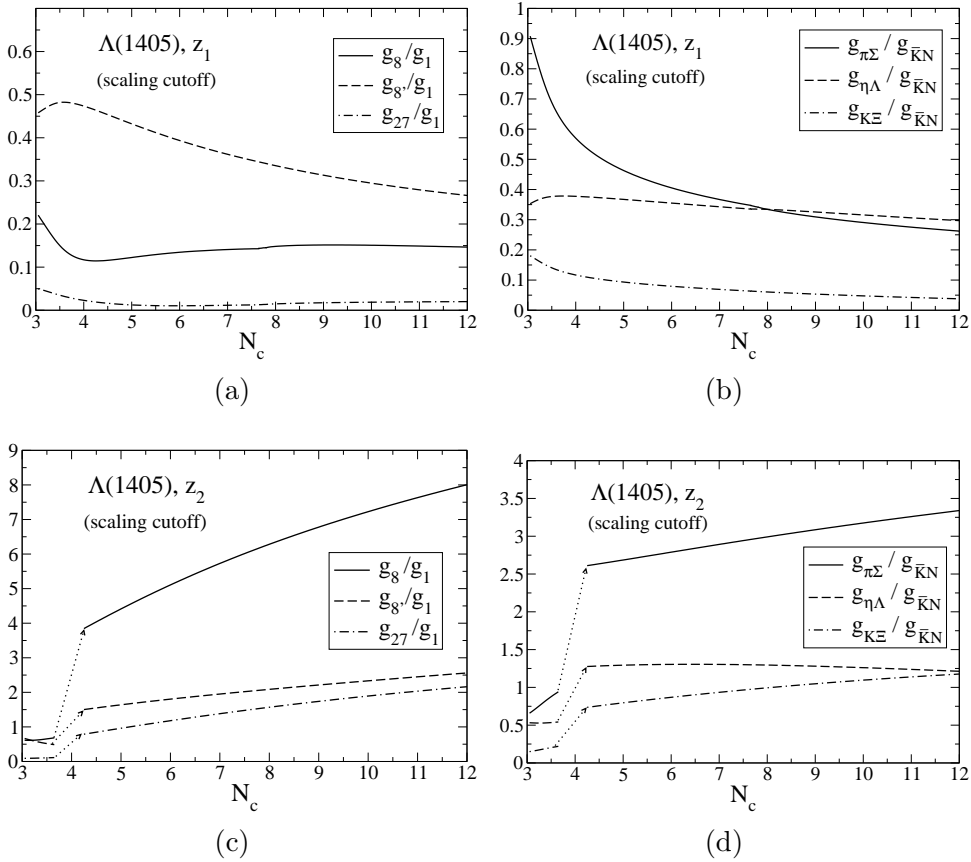

     \centering
     \subfigure[]{
          \label{fig:gcutoffa}
          \includegraphics[width=.45\linewidth]{figure5a.eps}}
       \subfigure[]{
          \label{fig:gcutoffb}
          \includegraphics[width=.45\linewidth]{figure5b.eps}}
   \\ \vspace{0.1cm}
       \subfigure[]{
          \label{fig:gcutoffc}
          \includegraphics[width=.45\linewidth]{figure5c.eps}}
       \subfigure[]{
          \label{fig:gcutoffd}
          \includegraphics[width=.45\linewidth]{figure5d.eps}}
     \caption{Couplings of 
 the $\Lambda(1405)$ poles using the scaling cutoff regularization
 method.} 
 \label{fig:gcutoff}
\end{figure}
Evaluating the residues in the same way, we see in
Fig.~\ref{fig:gcutoff} that the lower mass pole ($z_1$) is
dominated by the singlet  ($\bar{K}N$) in the scaling cutoff
scheme. Therefore, the dominance of the  flavor singlet and
$\bar{K}N$ components for the would-be-bound-state is  independent
of the choice of the regularization schemes and the  original 
position of the poles at $N_c=3$. This result leads us to
conjecture that  the pole showing the tendency to become a bound
state is smoothly connected to the  bound state found in the
idealized large $N_c$ limit.

\subsection{The $N_c$ scaling of the resonance parameters}

In this section, we investigate the internal structure of the resonances
based on the findings obtained by the above analyses.  
As explained in the introduction, the study of the $N_c$ dependence of the
resonance parameters, such as mass $M_R$,  excitation energy 
$\Delta E\equiv M_R-M-m$ and width $\Gamma_R$, can provide relevant 
information on the nature of the resonances by comparing to 
the quark picture.
QCD establishes particular $N_c$ behaviors for ordinary $qqq$ baryons as
\begin{equation}
    M_R  \sim {\cal O}(N_c) ,
    \quad
    \Delta E \sim{\cal O}(1) ,
    \quad
    \Gamma_R \sim {\cal O}(1) .
    \label{eq:scaling}
\end{equation}
Therefore, any deviation of our calculation from these general counting 
rules for $qqq$ baryons suggests  the  dynamical origin of the resonances under consideration.

\begin{figure}[tbp]
    \centering
    \includegraphics[width=.45\textwidth,clip]{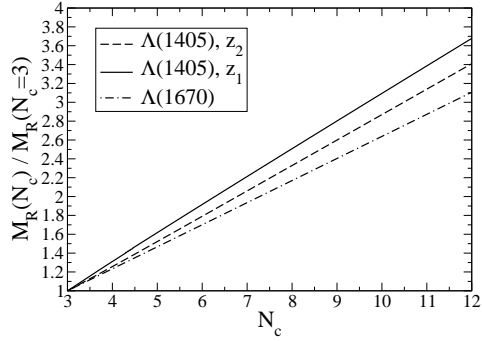}
 \caption{Resonance masses as functions of $N_c$ for the three
  poles normalized to the value at $N_c=3$. The calculation
 is done with the dimensional regularization method.}
 \label{fig:MR}
\end{figure}

Let us study first the $N_{c}$ dependence of the mass of the resonance. The
mass of the resonance is obtained by the real part of the pole position. 
In Fig.~\ref{fig:MR}, we show the masses of the resonances normalized by the
values at $N_c=3$, $M_R(N_c)/M_R(3)$, for the three poles obtained with the 
dimensional regularization method. 
This figure shows that an almost linear $N_c$ dependence is 
obtained for each state. This looks consistent with Eq.~\eqref{eq:scaling}, 
but it may not directly lead to the conclusion
 that these resonances are $qqq$ dominant 
states. For instance, the $N_c$ scaling of a meson-baryon molecule state 
would be $M_{MB}\sim M_B +m \sim {\cal O}(N_c)$ where we used the scaling of
ground state mesons and baryons, and assumed that the binding (excited) 
energy of the molecule is small compared to $M_{MB}$. If this is the case, the meson-baryon 
molecule state cannot be distinguished from the $qqq$ state, by looking at 
the $N_c$ behavior of the masses.
 In this respect, the $N_c$ scaling of the mass is not very 
useful to disentangle the $qqq$ state from the dynamical content.

\begin{figure}[tbp]
    \centering
    \includegraphics[width=.45\textwidth,clip]{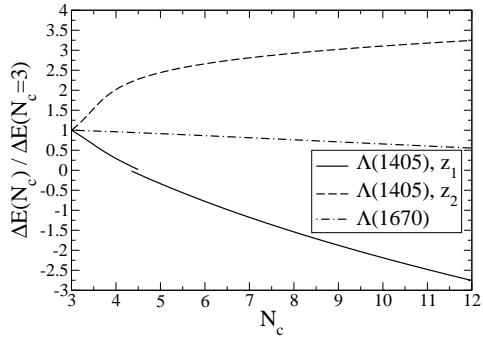}
 \caption{Resonance excitation energy, 
 ($\Delta E\equiv M_R-M_B-m$),
  as a function of $N_c$ for the three
 different poles normalized to the value at $N_c=3$. The calculation
 is done for the dimensional regularization method.}
 \label{fig:DM}
\end{figure}

Next we consider the $N_{c}$ scaling of the excitation energy.
In Fig.~\ref{fig:DM} we plot the excitation (or binding) energies
$\Delta E\equiv M_R-M_B-m$ of the three poles 
as functions of $N_c$. 
 We have normalized $\Delta E$ to the value
at
$N_c=3$ for each particular pole.
QCD 
predicts\footnote{Strictly speaking, the general counting 
rule tells us that
$M_R-M_B\sim {\cal O}(1)$. Combined with $m\sim {\cal O}(1)$,
we can infer the present definition of the excitation energy
$\Delta E\sim {\cal O}(1)$.} 
for $qqq$ baryons that
 $\Delta E\sim {\cal O}(1)$
\cite{Witten:1979kh,Goity:2004pw,Cohen:2003fv}. We see
in Fig.~\ref{fig:DM} that our results do not manifest the constant
behavior. For all the poles
 $\Delta E$ increases in their magnitude as $N_c$ increases.
Since $\Delta E$ is much smaller than $M_R$ and
 $M_B+m$, it barely modifies the linear behavior of $M_R$ in  
Fig.~\ref{fig:MR}. One should keep in mind  that these
resonances are generated within a coupled channel model and hence
several meson-baryon states ($\bar K N$, $\pi\Sigma$, 
$\eta\Lambda$ and $K\Xi$) contribute to their composition.
Therefore, the simplified interpretation of the binding
(excitation) energy should be taken with care. Hence, the study of
the $N_c$ dependence of $\Delta E$ should also be considered as
indicative but not conclusive.

 \begin{figure}[tbp]
    \centering
    \includegraphics[width=.45\textwidth,clip]{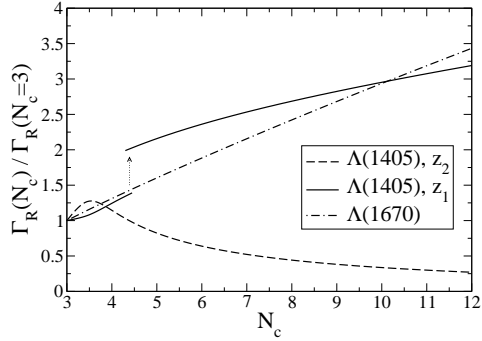}
 \caption{$N_c$ behavior of the widths of the two-$\Ls$ and the
 $\Lambda(1670)$ resonance. The calculation
 is done for the dimensional regularization method.}
 \label{fig:G}
\end{figure}%

Much more illustrative and conclusive statements for the  understanding of the 
nature of these resonances can be obtained by studying the
$N_{c}$ scaling of the resonance width, $\Gamma_R$. The resonance width is 
given by twice the imaginary part of the pole position.  
In Fig.~\ref{fig:G} we show $\Gamma_R$  normalized to the value at
$N_c=3$ for the dimensional regularization  method. The general QCD
counting rule for $qqq$ baryons predicts  ${\cal O}(1)$ behavior
for the width of an excited baryon. Our results are clearly
different from this constant behavior. The width of the higher
$\Ls$ pole (dashed line), which tends to
become a singlet state,  goes to zero  as $N_{c}$ increases. This is
again essentially a  consequence
 of the non-trivial $N_c$ dependence of the
meson-baryon  interaction steming
 from the $N_c$ dependence of the
$SU(3)$ representation for baryons.
 The widths of the other $\Ls$ and the
$\Lambda(1670)$ poles increase with $N_c$. In all the cases the
widths of the resonances do not follow the constant  behavior
expected from the general counting rules for the widths of $qqq$ 
states. We have checked that the results with the other
regularization  methods (cutoff), not shown in the figure, are similar and
hence our results are  independent of the choice of the
renormalization procedure and its  $N_c$ dependence. 


\section{Conclusions}

We have studied the $N_c$ behavior of the 
baryon resonances and meson-baryon
scatterings with $S=-1$ and $I=0$ 
from the viewpoint of the chiral unitary approach.
Introducing $N_c$ dependence of the model parameters, we have analyzed two 
different situations; meson-baryon scattering in the large $N_c$ limit, and 
the $N_c$ behavior of the $\Ls$ and $\Lambda(1670)$ resonances 
not very far from $N_{c}=3$.
We have discussed important and nontrivial $N_c$ dependence in the 
Weinberg-Tomozawa interaction, which leads to an ${{\cal O}(1)}$ attraction 
in the large $N_c$ limit for the flavor singlet channels in the $SU(3)$ 
bases. The attractive interaction in this channel is strong enough to create
a bound state, unlike the states associated to the other representations. 
The existence of this bound state does not depend on the framework of the 
large $N_c$ extension of baryon representations. We have obtained the 
coupling strengths in the isospin basis through the Clebsch-Gordan 
coefficients with $N_c$ dependence. It is found that the transition between 
isospin channels vanishes in the large $N_c$ limit, and the diagonal 
$\bar{K}N$ channel manifests an ${{\cal O}(1)}$ attraction, which is again 
sufficiently strong to generate a bound state in the large $N_c$ limit.

In order to investigate the properties of physical resonances, we have 
explicitly broken flavor $SU(3)$ symmetry by using the physical 
pseudoscalar and baryon masses. On top of the trivial $N_c$ dependences of 
hadron masses and decay constants, we have included the $N_c$ dependences of
the WT interaction and the cutoff scale in different regularization 
methods. The positions and residues of the poles associated to the $\Ls$ and the
$\Lambda(1670)$ resonances have been studied as functions of $N_c$. 
It is found that one of the two $\Ls$ poles tends to become a bound state, 
and the other $\Ls$ and the $\Lambda(1670)$ poles are 
dissolving into the scattering states. The coupling strengths tell us that 
the pole becoming the bound state is essentially dominated by the flavor 
singlet and isospin $\bar{K}N$ components, while the dissolving states are 
made of other components. This observation indicates that the pole with 
decreasing width will eventually become the bound state found in the large 
$N_c$ limit. These results are independent  of  the renormalization 
procedure and its $N_c$ dependence.  

We have also evaluated the $N_c$ dependence of the mass, excitation energy 
and width of the $\Ls$ and the $\Lambda(1670)$ resonances. The results for 
the width are at odds with the general QCD counting rules for $qqq$ states. 
This means that the nature of these resonances are definitely not dominated 
by the $qqq$ component. These findings reinforce the meson-baryon 
``molecule'' nature of the $\Ls$ and the $\Lambda(1670)$ resonances within 
the chiral unitary approach.

The technique used and the results obtained regarding the $N_c$ 
behavior of these baryonic resonances represent a step forward in the 
understanding of the connection with the underlying QCD degrees of freedom
and the method can be apllied to study other baryonic resonances.

\section*{Acknowledgments}

T.H. thanks Professor Masayasu Harada for his stimulating comment on the 
chiral unitary approach, which partly motivated this work. T.H. is grateful 
to Professor Micha{\l} Prasza{\l}owicz for the discussion on the different 
large $N_c$ extensions of baryons during the YKIS2006 on ``New Frontiers 
in QCD". T.~H. thanks the Japan Society for the Promotion of Science (JSPS) 
for financial support. This work is supported in part by the Grant for 
Scientific Research (No.\ 19853500 and No.\ 18042001). L.~R. thanks 
financial support from MEC (Spain) grants No. FPA2004-03470, FIS2006-03438, 
FPA2007-62777, Fundaci\'on S\'eneca grant No. 02975/PI/05, European Union 
grant No. RII3-CT-20004-506078 and the Japan(JSPS)-Spain collaboration 
agreement. A part of this research was done under Yukawa International Program for 
Quark-Hadron Sciences.





\begin{thebibliography}{99}

\bibitem{Jaffe:1976ig}
 R.~L.~Jaffe,
 Phys.\ Rev.\  D 15 (1977) 267.
 
\bibitem{Alford:2000mm}
  M.~G.~Alford and R.~L.~Jaffe,
  Nucl.\ Phys.\  B 578 (2000) 367.

\bibitem{Kunihiro:2003yj}
  T.~Kunihiro, S.~Muroya, A.~Nakamura, C.~Nonaka, M.~Sekiguchi and H.~Wada
  [SCALAR Collaboration],
  Phys.\ Rev.\  D 70 (2004) 034504.
  
\bibitem{McNeile:2007fu}
  C.~McNeile,
  arXiv:0710.0985 [hep-lat].

\bibitem{Dobado:1996ps}
  A.~Dobado and J.~R.~Pelaez,
  Phys.\ Rev.\  D 56 (1997) 3057.

\bibitem{Oller:1997ti}
  J.~A.~Oller and E.~Oset,
  Nucl.\ Phys.\  A 620 (1997) 438,
  Erratum-ibid.\  A 652 (1999) 407.
  
\bibitem{Oller:1998hw}
  J.~A.~Oller, E.~Oset and J.~R.~Pelaez,
  Phys.\ Rev.\  D 59 (1999) 074001,
  Erratum-ibid.\  D 60 (1999) 099906 , D75 (2007) 099903.

\bibitem{Oller:1999zr}
J.~A.~Oller and E.~Oset,
Phys.\ Rev.\ D 60 (1999) 074023.

\bibitem{Dalitz:1960du}
  R.~H.~Dalitz and S.~F.~Tuan,
  Annals Phys.\ 10 (1960) 307. 

\bibitem{Dalitz:1967fp}
  R.~H.~Dalitz, T.~C.~Wong and G.~Rajasekaran,
  Phys.\ Rev.\ 153 (1967) 1617.
  
\bibitem{Isgur:1978xj}
  N.~Isgur and G.~Karl,
  Phys.\ Rev.\  D 18 (1978) 4187.

\bibitem{Melnitchouk:2002eg}
  W.~Melnitchouk {\it et al.},
  Phys.\ Rev.\  D 67 (2003) 114506.
  
\bibitem{Nemoto:2003ft}
  Y.~Nemoto, N.~Nakajima, H.~Matsufuru and H.~Suganuma,
  Phys.\ Rev.\  D 68 (2003) 094505.

\bibitem{Lee:2005mr}
  F.~X.~Lee and C.~Bennhold,
  Nucl.\ Phys.\  A 754 (2005) 248.

\bibitem{Ishii:2007ym}
  N.~Ishii, T.~Doi, M.~Oka and H.~Suganuma,
  Prog.\ Theor.\ Phys.\ Suppl.\  168 (2007) 598.
  
\bibitem{Kaiser:1995eg}
N.~Kaiser, P.~B. Siegel, and W.~Weise,
\newblock Nucl. Phys. A 594 (1995) 325.

\bibitem{Oset:1998it}
E.~Oset and A.~Ramos,
\newblock Nucl. Phys. A 635 (1998) 99.
  
\bibitem{Oller:2000fj}
  J.~A.~Oller and U.~G.~Meissner,
  Phys.\ Lett.\ B 500 (2001) 263.

\bibitem{Lutz:2001yb}
M.~F.~M. Lutz and E.~E. Kolomeitsev,
\newblock Nucl. Phys. A 700 (2002) 193.

\bibitem{Hyodo:2007np}
  T.~Hyodo, D.~Jido and L.~Roca,
  \newblock Phys. Rev. D 77 (2008) 056010.

\bibitem{Weinberg:1979kz}
S.~Weinberg,
\newblock Physica A 96 (1979) 327.
    
\bibitem{Gasser:1983yg}
J.~Gasser and H.~Leutwyler,
Annals Phys.\ 158 (1984) 142.

\bibitem{Gasser:1984gg}
J.~Gasser and H.~Leutwyler,
Nucl.\ Phys.\ B 250 (1985) 465.

\bibitem{Kaiser:1998fi}
  N.~Kaiser,
  Eur.\ Phys.\ J.\  A 3 (1998) 307.

\bibitem{Markushin:2000fa}
  V.~E.~Markushin,
  Eur.\ Phys.\ J.\  A 8 (2000) 389.

\bibitem{Lutz:2003fm}
  M.~F.~M.~Lutz and E.~E.~Kolomeitsev,
  Nucl.\ Phys.\  A 730 (2004) 392.

\bibitem{Roca:2005nm}
  L.~Roca, E.~Oset and J.~Singh,
  Phys.\ Rev.\  D 72 (2005) 014002.
  
\bibitem{Kaiser:1995cy}
  N.~Kaiser, P.~B.~Siegel and W.~Weise,
  Phys.\ Lett.\  B 362 (1995) 23.

\bibitem{Kaiser:1996js}
  N.~Kaiser, T.~Waas and W.~Weise,
  Nucl.\ Phys.\  A 612 (1997) 297.
  
\bibitem{Krippa:1998us}
  B.~Krippa,
  Phys.\ Rev.\  C 58 (1998) 1333.
  
\bibitem{Nieves:1998hp}
  J.~Nieves and E.~Ruiz Arriola,
  Phys.\ Lett.\  B 455 (1999) 30.

\bibitem{Nacher:1999vg}
  J.~C.~Nacher, A.~Parreno, E.~Oset, A.~Ramos, A.~Hosaka and M.~Oka,
  Nucl.\ Phys.\  A 678 (2000) 187.

\bibitem{Oset:2001cn}
  E.~Oset, A.~Ramos and C.~Bennhold,
  Phys.\ Lett.\ B 527 (2002) 99,
  Erratum-ibid.\ B 530 (2002) 260.

\bibitem{Inoue:2001ip}
  T.~Inoue, E.~Oset and M.~J.~Vicente Vacas,
  Phys.\ Rev.\  C 65 (2002) 035204.

\bibitem{Jido:2002yz}
  D.~Jido, A.~Hosaka, J.~C.~Nacher, E.~Oset and A.~Ramos,
  Phys.\ Rev.\  C {\bf 66} (2002) 025203
  
\bibitem{Jido:2002zk}
  D.~Jido, E.~Oset and A.~Ramos,
  Phys.\ Rev.\  C {\bf 66} (2002) 055203

\bibitem{Weinberg:1966kf}
  S.~Weinberg,
  Phys.\ Rev.\ Lett.\  17 (1966) 616.

\bibitem{Tomozawa:1966jm}
  Y.~Tomozawa,
  Nuovo Cim.\  46A (1966) 707.
  
\bibitem{Hyodo:2006yk}
T.~Hyodo, D.~Jido, and A.~Hosaka,
\newblock Phys. Rev. Lett. 97 (2006) 192002.

\bibitem{Hyodo:2006kg}
T.~Hyodo, D.~Jido, and A.~Hosaka,
\newblock Phys. Rev. D 75 (2007) 034002.

\bibitem{Fink:1989uk}
  P.~J.~Fink, G.~He, R.~H.~Landau and J.~W.~Schnick,
  Phys.\ Rev.\  C {\bf 41} (1990) 2720.
  
\bibitem{Jido:2003cb}
D.~Jido, J.~A. Oller, E.~Oset, A.~Ramos, and U.~G. Meissner,
\newblock Nucl. Phys. A 725 (2003) 181.



\bibitem{Hyodo:2002pk}
  T.~Hyodo, S.~I.~Nam, D.~Jido and A.~Hosaka,
  Phys.\ Rev.\  C 68 (2003) 018201.
  
\bibitem{Hyodo:2003qa}
T.~Hyodo, S.~I. Nam, D.~Jido, and A.~Hosaka,
\newblock Prog. Theor. Phys. 112 (2004) 73.

\bibitem{Borasoy:2005ie}
 B.~Borasoy, R.~Nissler, and W.~Weise,
 \newblock Eur. Phys. J. A 25 (2005) 79.

\bibitem{Oller:2005ig}
 J.~A. Oller, J.~Prades, and M.~Verbeni,
 \newblock Phys. Rev. Lett. 95 (2005) 172502.

\bibitem{Oller:2006jw}
 J.~A. Oller,
 \newblock Eur. Phys. J. A 28 (2006) 63.

\bibitem{Borasoy:2006sr}
 B.~Borasoy, U.~G. Meissner, and R.~Nissler,
 \newblock Phys. Rev. C 74 (2006) 055201.

\bibitem{Hyodo:2007jq}
 T.~Hyodo and W.~Weise,
 \newblock Phys. Rev. C 77 (2008) 03524. 

\bibitem{Prakhov:2004an}
  S.~Prakhov {\it et al.} [Crystall Ball Collaboration],
  Phys.\ Rev.\ C 70 (2004) 034605.

\bibitem{Magas:2005vu}
V.~K. Magas, E.~Oset, and A.~Ramos,
\newblock Phys. Rev. Lett. 95 (2005) 052301.


\bibitem{Zychor:2007gf}
  I.~Zychor {\it et al.},
  Phys.\ Lett.\  B {\bf 660} (2008) 167.

\bibitem{Geng:2007vm}
  L.~S.~Geng and E.~Oset,
  arXiv:0707.3343 [hep-ph].

\bibitem{Hyodo:2003jw}
T.~Hyodo, A.~Hosaka, E.~Oset, A.~Ramos, and M.~J. Vicente~Vacas,
\newblock Phys. Rev. C 68 (2003) 065203.

\bibitem{Geng:2007hz}
  L.~S.~Geng, E.~Oset and M.~Doring,
  Eur.\ Phys.\ J.\  A 32 (2007) 201.

\bibitem{Hyodo:2004vt}
T.~Hyodo, A.~Hosaka, M.~J. Vicente~Vacas, and E.~Oset,
\newblock Phys. Lett. B 593 (2004) 75.
  
\bibitem{Hemingway:1984pz}
  R.~J.~Hemingway,
  Nucl.\ Phys.\  B {\bf 253} (1985) 742.

\bibitem{Dalitz:1991sq}
  R.~H.~Dalitz and A.~Deloff,
  J.\ Phys.\ G {\bf 17} (1991) 289.
 
  
  
\bibitem{Akaishi:2002bg}
  Y.~Akaishi and T.~Yamazaki,
  Phys.\ Rev.\  C 65 (2002) 044005.
 
\bibitem{Yamazaki:2007cs}
  T.~Yamazaki and Y.~Akaishi,
  Phys.\ Rev.\  C 76 (2007) 045201.

\bibitem{Dote:2008in}
  A.~Dot\'{e}, T.~Hyodo and W.~Weise,
  arXiv:0802.0238 [nucl-th], Nucl. Phys. A, in press.
  
\bibitem{Kaplan:1986yq}
  D.~B.~Kaplan and A.~E.~Nelson,
  Phys.\ Lett.\  B 175 (1986) 57.
  
\bibitem{Hooft:1974jz}
G.~'t Hooft,
Nucl.\ Phys.\ B 72 (1974) 461.

\bibitem{Witten:1979kh}
E.~Witten,
\newblock Nucl. Phys. B 160 (1979) 57.

\bibitem{Witten:1980sp}
E.~Witten,
Annals Phys.\  128 (1980) 363.

\bibitem{Pelaez:2003dy}
  J.~R.~Pelaez,
  Phys.\ Rev.\ Lett.\  92 (2004) 102001.

\bibitem{Pelaez:2006nj}
  J.~R.~Pelaez and G.~Rios,
  Phys.\ Rev.\ Lett.\  97 (2006) 242002.

\bibitem{Pelaez:2004xp}
  J.~R.~Pelaez,
  Mod.\ Phys.\ Lett.\ A 19 (2004) 2879.

\bibitem{newaxialsNc}
L.~S. Geng, E.~Oset, J.~R.~Pelaez ad L.~Roca, {\it 
submitted to Phys. Lett. B
}, preprint IFIC-07-1028

\bibitem{Jaffe:2007id}
  R.~L.~Jaffe,
  Prog.\ Theor.\ Phys.\ Suppl.\  {\bf 168} (2007) 127.

\bibitem{Karl:1985qy}
G.~Karl, J.~Patera, and S.~Perantonis,
\newblock Phys. Lett. B 172 (1986) 49.

\bibitem{Dulinski:1987er}
Z.~Dulinski and M.~Praszalowicz,
\newblock Acta Phys. Polon. B 18 (1988) 1157.

\bibitem{Dulinski:1988yh}
Z.~Dulinski,
\newblock Acta Phys. Polon. B 19 (1988) 891.

\bibitem{Callan:1985hy}
C.~G. Callan and I.~R. Klebanov,
\newblock Nucl. Phys. B 262 (1985) 365.

\bibitem{Itzhaki:2003nr}
N.~Itzhaki, I.~R. Klebanov, P.~Ouyang, and L.~Rastelli,
\newblock Nucl. Phys. B 684 (2004) 264.

\bibitem{Goity:2004pw}
 J.~L. Goity,
 \newblock Phys. Atom. Nucl. 68 (2005) 624.

\bibitem{Cohen:2003fv}
 T.~D. Cohen, D.~C. Dakin, A.~Nellore, and R.~F. Lebed,
 \newblock Phys. Rev. D 69 (2004) 056001.

\bibitem{GarciaRecio:2006wb}
  C.~Garcia-Recio, J.~Nieves and L.~L.~Salcedo,
  Phys.\ Rev.\  D 74 (2006) 036004.

\bibitem{Bernard:1995dp}
  V.~Bernard, N.~Kaiser and U.~G.~Meissner,
  Int.\ J.\ Mod.\ Phys.\  E 4 (1995) 193.

\bibitem{deSwart:1963gc}
J.~J. de~Swart,
\newblock Rev. Mod. Phys. 35 (1963) 916.

\bibitem{McNamee:1964xq}
  P.~S.~J.~McNamee and F.~Chilton,
  Rev.\ Mod.\ Phys.\  36 (1964) 1005.

\bibitem{Jenkins:1995td}
  E.~E.~Jenkins and R.~F.~Lebed,
  Phys.\ Rev.\  D {\bf 52} (1995) 282
  
\bibitem{Dashen:1993as}
  R.~F.~Dashen and A.~V.~Manohar,
  Phys.\ Lett.\  B {\bf 315} (1993) 425.

\bibitem{Dashen:1993ac}
  R.~F.~Dashen and A.~V.~Manohar,
  Phys.\ Lett.\  B {\bf 315} (1993) 438.
  
    

\bibitem{Cohen:2004ki}
T.~D. Cohen and R.~F. Lebed,
\newblock Phys. Rev. D 70 (2004) 096015.
  
\bibitem{Hyodo:2007jk}
  T.~Hyodo, D.~Jido and A.~Hosaka,
  Prog.\ Theor.\ Phys.\ Suppl.\  168 (2007) 32.

\bibitem{Hyodo:2008xr}
  T.~Hyodo, D.~Jido and A.~Hosaka,
  arXiv:0803.2550 [nucl-th].
  
\bibitem{Piesciuk:2007xq}
  K.~Piesciuk and M.~Praszalowicz,
  Prog.\ Theor.\ Phys.\ Suppl.\  168 (2007) 70.
  
\bibitem{shadowpoles}
R.~J.~Eden and J.~R.~Taylor,
  Phys.\ Rev.\ 133, B1575 - B1580 (1964);
M.~Ross, Phys.\ Rev.\ Lett.\ 11, 450 - 453 (1963).


\end{thebibliography}
\end{document}